\documentclass[onecolumn,pra,nofootinbib]{revtex4-2}

\usepackage{graphicx} 
\usepackage{amsfonts}
\usepackage{amssymb}
\usepackage{amscd}
\usepackage{amsmath}
\usepackage{enumerate}
\usepackage{epsfig}
\usepackage{subfigure}
\usepackage{subfloat}
\usepackage{xcolor}
\usepackage{amsthm}
\usepackage{physics}
\usepackage[most]{tcolorbox}
\usepackage{ulem}
\usepackage{tikz}
\usetikzlibrary{quantikz}
\usetikzlibrary{backgrounds}
\usetikzlibrary{shapes.geometric}
\usepackage{tabularx}
\usepackage{float}
\usepackage{appendix}
\usepackage{makecell}
\usepackage[ruled,vlined]{algorithm2e}
\usepackage{algpseudocode}
\usepackage{dsfont}
\usepackage[colorlinks, linkcolor=red, anchorcolor=blue, citecolor=green]{hyperref}
\usepackage{MnSymbol}
\usepackage{pgfplots}
\pgfplotsset{compat=1.18}
\makeatletter
\newcommand{\ignorelegend}[2][]{}
\makeatother

\newtcolorbox[auto counter]{mybox}[2]{
	enhanced,
	breakable,
	label=#1,
	colback=blue!5!white,
	colframe=blue!75!black,
	fonttitle=\bfseries,
	title=Box \thetcbcounter: #2
}

\newtheorem{lemma}{Lemma}

\newcommand{\lket}[1]{\vert #1 \rangle\!\rangle}
\newcommand{\lbra}[1]{\langle\!\langle #1 \vert}
\newcommand{\lbraket}[2]{\langle\!\langle #1 \vert #2 \rangle\!\rangle}
\newcommand{\lketbra}[2]{\vert #1 \rangle\!\rangle\langle\!\langle #2 \vert}

\newcommand{\tra}[1]{\mathrm{tr}(#1)}

\newcommand{\comments}[1]{}

\begin{document}

\title{Spectral tomography and echo benchmarking for arbitrary two-qubit gates}
\author{Han Ye}
\author{Guoding Liu}
\author{Xiongfeng Ma}
\email{xma@tsinghua.edu.cn}
\affiliation{Center for Quantum Information, Institute for Interdisciplinary Information Sciences, Tsinghua University, Beijing, 100084 China}

\begin{abstract}
Characterizing arbitrary quantum gates robustly against state preparation and measurement errors remains challenging, particularly beyond standard gate sets such as the Clifford group. A strategy is to exploit the algebraic structure of quantum gates under local unitary transformations, where gates reduce to canonical forms with inherent symmetries. For arbitrary two-qubit gates, the KAK decomposition yields a canonical form whose dynamics naturally decouple into four small subspaces. Building on this structure with only ideal local auxiliary operations, we construct two complementary protocols that remain robust to state preparation and measurement errors. The first protocol isolates each subspace to simplify spectral reconstruction. The second constructs echo circuits to cancel the ideal evolution within the measured subspace and estimates gate fidelity from the resulting single-exponential decays. The two protocols can be used independently or combined, where spectral information helps improve fidelity estimation by correcting coherent terms missed by the echo decays. Numerical simulations demonstrate accurate spectral reconstruction and substantial improvements in fidelity estimation, illustrating how canonical gate structures can guide general and efficient quantum benchmarking.
\end{abstract}

\maketitle

\section{Introduction}\label{sec: intro}
Accurate quantum gate characterization is essential for reliable quantum computing, providing the information needed for performance assessment, error diagnosis, and control calibration~\cite{PhysRevLett.112.240504,Blume_Kohout_2017,Fan_2025}. Two established approaches are quantum process tomography (QPT) and randomized benchmarking (RB). Standard QPT reconstructs a complete description of the implemented channels, but typically requires substantial resources and relies on accurately characterized state preparation and measurement (SPAM) operations~\cite{PhysRevLett.78.390,Chuang_1997}. Unaccounted SPAM errors can therefore bias its reconstruction. RB instead estimates only a few gate-quality metrics, such as average fidelity, generally using fewer resources and providing robustness to SPAM errors~\cite{PhysRevA.77.012307,PhysRevLett.106.180504}. These approaches offer different trade-offs between the amount of information obtained and the resources required.

Multi-qubit gates are fundamental building blocks of quantum computation. In particular, any entangling two-qubit gate, together with arbitrary single-qubit gates, suffices for universality~\cite{Bremner_2002}. Physical platforms increasingly support diverse families of native entangling operations beyond Clifford gates, including iSWAP-like, controlled-phase, and continuously parameterized fSim gates in superconducting processors~\cite{PhysRevLett.127.180501,PhysRevLett.123.210501,Foxen_2020}, M{\o}lmer--S{\o}rensen gates with tunable entangling angles in trapped-ion processors~\cite{PhysRevLett.117.060505,Yale_2025}, and CCZ gates in neutral atom processors~\cite{Evered_2023}. Recent control schemes can also implement arbitrary two-qubit gates~\cite{chen2025efficientimplementationarbitrarytwoqubit}. On the other hand, multi-qubit gates still dominate the overall error budget in current quantum processors~\cite{eickbusch2026resolvingsuppressingtemporallycorrelated}. Accurately characterizing and optimizing this broad range of operations is therefore critical for improving circuit performance and achieving scalable quantum computation. However, SPAM-robust characterization of arbitrary multi-qubit gates remains challenging, particularly for non-Clifford gates. The applicability of a method depends on its resource costs and on the assumptions it makes about the target gate, noise, and auxiliary operations.

Existing protocols address different aspects of this challenge. On the QPT side, gate set tomography (GST) jointly reconstructs descriptions of gates and SPAM, providing detailed, self-consistent characterization without requiring pre-calibrated SPAM operations~\cite{blumekohout2013robustselfconsistentclosedformtomography,Nielsen_2021}. Spectral quantum tomography (SQT) extracts noisy-channel eigenvalues from repeated-gate signals in a SPAM-resistant manner~\cite{Helsen_2019}. Based on SQT, channel spectrum benchmarking (CSB) uses knowledge of the ideal target to probe selected eigenvalues~\cite{Gu_2023}. On the RB side, methods for Clifford gates are well established~\cite{PhysRevLett.106.180504,Magesan_2012,Erhard_2019}, and subsequent extensions have broadened benchmarking to various classes of non-Clifford operations~\cite{Cross_2016,Ye_2026}. These approaches offer different balances between the information obtained, the auxiliary operations used, the assumptions accepted, and the resources required. Across arbitrary multi-qubit non-Clifford operations, achieving SPAM-robust characterization while keeping experimental and computational costs manageable remains a practical challenge.

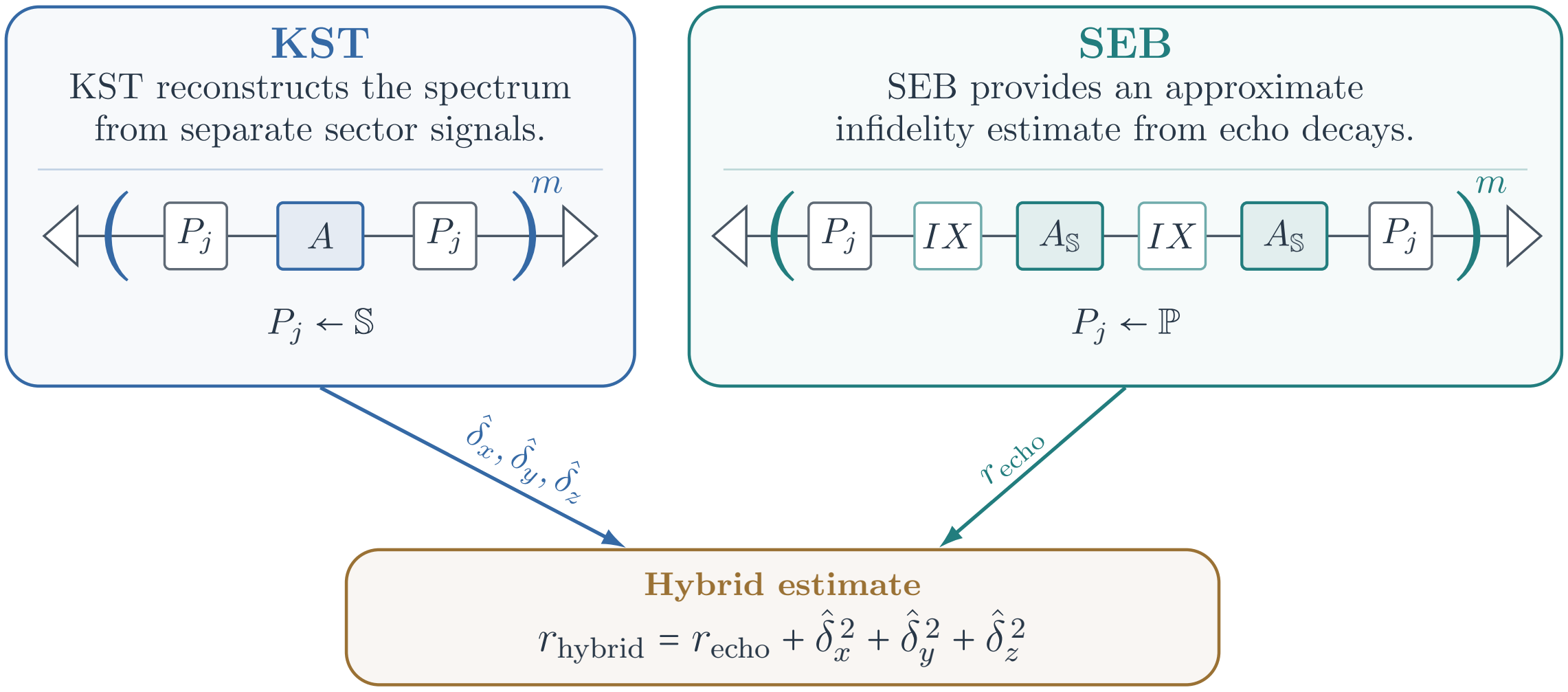
\begin{figure}[h!]
    \centering
    \definecolor{KSTBlue}{HTML}{3569A5}
    \definecolor{SEBTeal}{HTML}{227D7E}
    \definecolor{HybridGold}{HTML}{9A7134}
    \definecolor{DiagramInk}{HTML}{283747}
    \begin{tikzpicture}[
        x=1cm,y=1cm,
        font=\normalsize,text=DiagramInk,
        panel/.style={rectangle,rounded corners=9pt,line width=0.85pt,inner sep=0pt},
        gate/.style={draw=DiagramInk!75,fill=white,rounded corners=1.5pt,line width=0.65pt,minimum width=0.60cm,minimum height=0.64cm,inner sep=2pt},
        target/.style={gate,minimum width=0.82cm,line width=0.8pt},
        wire/.style={draw=DiagramInk!85,line width=0.65pt},
        flow/.style={line width=1.0pt,-{Latex[length=2.4mm,width=1.6mm]}},
        explanation/.style={font=\normalsize,align=center,inner sep=0pt},
        flow label/.style={font=\normalsize,sloped,above,fill=white,inner sep=2pt}
        ]

        \node[panel,draw=KSTBlue,fill=KSTBlue!4,minimum width=6.05cm,minimum height=3.65cm] (kst) at (0,0.10) {};
        \node[panel,draw=SEBTeal,fill=SEBTeal!4,minimum width=8.40cm,minimum height=3.65cm] (seb) at (7.75,0.10) {};
        \node[font=\large\bfseries,text=KSTBlue] at (0,1.58) {KST};
        \node[font=\large\bfseries,text=SEBTeal] at (7.75,1.58) {SEB};
        \node[explanation,text width=5.40cm] at (0,0.93) {KST reconstructs the spectrum from separate sector signals.};
        \node[explanation,text width=7.65cm] at (7.75,0.93) {SEB provides an approximate\\infidelity estimate from echo decays.};
        \draw[draw=KSTBlue!30,line width=0.5pt] (-2.72,0.36) -- (2.72,0.36);
        \draw[draw=SEBTeal!30,line width=0.5pt] (3.88,0.36) -- (11.62,0.36);

        \draw[wire] (-2.34,-0.28) -- (2.34,-0.28);
        \draw[wire,fill=white] (-2.66,-0.28) -- (-2.34,0.00) -- (-2.34,-0.56) -- cycle;
        \draw[wire,fill=white] (2.66,-0.28) -- (2.34,0.00) -- (2.34,-0.56) -- cycle;
        \node[font=\Huge,text=KSTBlue] at (-1.97,-0.28) {$(\vphantom{A}$};
        \node[font=\Huge,text=KSTBlue] at (1.97,-0.28) {$)\vphantom{A}$};
        \node[anchor=south west,inner sep=0pt,text=KSTBlue] at (2.02,0.13) {$m$};
        \node[gate] at (-1.20,-0.28) {$P_j$};
        \node[target,draw=KSTBlue,fill=KSTBlue!13] at (0,-0.28) {$A$};
        \node[gate] at (1.20,-0.28) {$P_j$};
        \node at (0,-1.15) {$P_j\leftarrow\mathbb S$};

        \draw[wire] (4.10,-0.28) -- (11.43,-0.28);
        \draw[wire,fill=white] (3.78,-0.28) -- (4.10,0.00) -- (4.10,-0.56) -- cycle;
        \draw[wire,fill=white] (11.75,-0.28) -- (11.43,0.00) -- (11.43,-0.56) -- cycle;
        \node[font=\Huge,text=SEBTeal] at (4.44,-0.28) {$(\vphantom{A}$};
        \node[font=\Huge,text=SEBTeal] at (11.06,-0.28) {$)\vphantom{A}$};
        \node[anchor=south west,inner sep=0pt,text=SEBTeal] at (11.11,0.13) {$m$};
        \node[gate] at (5.00,-0.28) {$P_j$};
        \node[gate,draw=SEBTeal!65] at (6.04,-0.28) {$IX$};
        \node[target,draw=SEBTeal,fill=SEBTeal!13] at (7.12,-0.28) {$A_{\mathbb S}$};
        \node[gate,draw=SEBTeal!65] at (8.20,-0.28) {$IX$};
        \node[target,draw=SEBTeal,fill=SEBTeal!13] at (9.28,-0.28) {$A_{\mathbb S}$};
        \node[gate] at (10.40,-0.28) {$P_j$};
        \node at (7.75,-1.15) {$P_j\leftarrow\mathbb P$};

        \node[panel,draw=HybridGold,fill=HybridGold!6,minimum width=8.40cm,minimum height=1.30cm] (hybrid) at (4.45,-3.95) {};
        \node[font=\bfseries,text=HybridGold] at (4.45,-3.66) {Hybrid estimate};
        \node[font=\large] at (4.45,-4.15) {$r_{\mathrm{hybrid}}=r_{\mathrm{echo}}+\hat{\delta}_x^{\,2}+\hat{\delta}_y^{\,2}+\hat{\delta}_z^{\,2}$};
        \draw[flow,draw=KSTBlue] (kst.south) -- node[flow label,text=KSTBlue,pos=0.63] {$\hat{\delta}_x,\hat{\delta}_y,\hat{\delta}_z$} ([xshift=-1.50cm]hybrid.north);
        \draw[flow,draw=SEBTeal] (seb.south) -- node[flow label,text=SEBTeal,pos=0.53] {$r_{\mathrm{echo}}$} ([xshift=1.50cm]hybrid.north);
    \end{tikzpicture}
    \caption{Schematic representation of our characterization framework for a two-qubit target gate with KAK decomposition $U=LAR$, where $L$ and $R$ are local unitaries and $A=\exp[-i(c_xXX+c_yYY+c_zZZ)]$. At repetition $j$, KST samples $P_j$ uniformly from $\mathbb S=\{II,XX,YY,ZZ\}$ and SEB samples $P_j$ uniformly from the full two-qubit Pauli group $\mathbb P$. Each $A_{\mathbb S}$ denotes one use of the implemented $A$ twirled by $\mathbb S$. The two protocols can not only be run independently, but also provide complementary information to obtain the combined infidelity estimate $r_\textrm{hybrid}$.}
    \label{fig: schematic}
\end{figure}

A strategy for characterizing arbitrary gates is to exploit their algebraic structure under local unitary transformations, which reduces complex operations to canonical forms with inherent symmetries. In this work, we develop this approach for arbitrary two-qubit gates, which represent both a primary source of hardware errors~\cite{eickbusch2026resolvingsuppressingtemporallycorrelated} and a setting with a common algebraic structure given by the KAK (also called Cartan) decomposition~\cite{PhysRevA.67.042313,tucci2005introductioncartanskakdecomposition}. The KAK decomposition separates local single-qubit rotations from a canonical nonlocal core whose dynamics naturally decouple into four four-dimensional subspaces. Restricting all auxiliary operations to tensor products of single-qubit gates to avoid the overhead of additional entangling gates, and assuming these local auxiliary gates are ideal, we translate this structure into two complementary, SPAM-robust protocols. Building on SQT, KAK spectral tomography (KST) exploits the decoupled structure to obtain four separate spectral signals, each containing at most four eigenvalues, simplifying spectral reconstruction. Sector echo benchmarking (SEB) uses local echo circuits to cancel the ideal evolution within the measured subspace, producing single-exponential decays that yield an approximate fidelity. We further combine the spectral information from KST with the decay information from SEB to improve fidelity estimation. We derive the small-error relation underlying this combination and numerically demonstrate both the individual protocols and their combined use. The primary contributions of our work are summarized in Fig.~\ref{fig: schematic}.

The structure of this article is as follows. In Sec.~\ref{sec: preliminary}, we provide the necessary background. In Sec.~\ref{sec: kst}, we introduce the KST protocol. In Sec.~\ref{sec:seb}, we introduce the SEB protocol, and explain how to combine it with the KST protocol. In Sec.~\ref{sec: simulation}, we present numerical simulation experiments related to our protocols. Finally, we compare our protocols with existing methods in Sec.~\ref{sec: comparison} and conclude in Sec.~\ref{sec: discussion}.

\section{Preliminaries}\label{sec: preliminary}

\subsection{Pauli-Liouville representation}\label{subsec: pauli representation}

We consider quantum channels acting on an $n$-qubit system with Hilbert-space dimension $d=2^n$ (in this article $n$ is usually two). Unless otherwise stated, a quantum channel is a completely positive and trace-preserving (CPTP) map. The $n$-qubit Pauli group and the normalized Pauli operators, with global phases in $\{\pm1,\pm i\}$ identified, are
\begin{equation}
    \begin{gathered}
        \mathbb{P}=\{I,X,Y,Z\}^{\otimes n},\\
        \mathbb{P}'=\{\sigma_P=\frac{P}{\sqrt{d}}:P\in\mathbb{P}\},
    \end{gathered}
\end{equation}
with $\tra{\sigma_P^\dag\sigma_Q}=\delta_{PQ}$. Because the normalized Pauli operators form an orthonormal basis, any operator $O$ can be expanded in this basis and vectorized as
\begin{equation}
    \begin{aligned}
        O&=\sum_{P\in\mathbb{P}}\tra{\sigma_P^\dag O}\sigma_P\\
        \iff\lket{O}&=\sum_{P\in\mathbb{P}}\lbraket{\sigma_P}{O}\lket{\sigma_P}.
    \end{aligned}
\end{equation}
Here $\lbraket{A}{B}=\tra{A^\dag B}$ defines the inner product, $\{\lket{\sigma_P}\}$ forms an orthonormal basis under this inner product, and $\lket{O}$ is a $d^2$-dimensional vector. This representation is called the Pauli-Liouville representation.

Furthermore, a quantum channel $\Lambda$ can be represented by its Pauli transfer matrix (PTM),
\begin{equation}
    \Lambda_{PQ}=\lbraket{\sigma_P}{\Lambda(\sigma_Q)},
\end{equation}
where $P$ and $Q$ label the rows and columns respectively. The action and composition of channels then correspond to matrix multiplication.
\begin{equation}
    \begin{gathered}
        \lket{\Lambda(O)}=\Lambda\lket{O},\\
        \Lambda_2\circ\Lambda_1\longleftrightarrow\Lambda_2\Lambda_1.
    \end{gathered}
\end{equation}
Here we use the same symbol for a channel and its Pauli transfer matrix, with the meaning determined by context.

A valid quantum channel $\Lambda$ should be CPTP, which gives some properties of the PTM. Although ``complete positivity" is not directly apparent from the PTM, Hermiticity preservation implies that the PTM is real. With the identity Pauli placed first, ``trace preservation" yields the block form
\begin{equation}\label{eq: block structure}
    \Lambda=\begin{pmatrix}
        1 & \mathbf{0} \\
        t & T
    \end{pmatrix},
\end{equation}
where $T$ acts on the $(d^2-1)$-dimensional traceless operator subspace. The vector $t$ vanishes if and only if the channel is unital, while we allow a channel to be non-unital.

In this article, we denote unitaries by ordinary letters, such as $U$, $P$ and $Q$, and groups of unitaries by blackboard-bold letters, such as $\mathbb{P}$ and $\mathbb{S}$. A unitary $U$ induces the channel $\mathcal{U}(\rho)=U\rho U^\dagger$, represented by the corresponding calligraphic font. A tilde indicates a noisy implementation, such as $\tilde{\mathcal{U}}$.

\subsection{Characterization quantities}\label{subsec: characterization quantities}

Fidelity is a widely used metric to estimate the agreement between an implemented operation and its ideal target, while the spectrum of a PTM provides a different description of the channel. For example, when the correspondence between the ideal eigenvalues and the estimated eigenvalues can be established, phase deviations can indicate coherent errors, while eigenvalue moduli can quantify incoherent decays. Fidelity and spectral information thus offer different tools for characterization and calibration.

Since a PTM is real, its complex eigenvalues occur in complex-conjugate pairs, with equal algebraic multiplicities. All the eigenvalues of a valid PTM have moduli no larger than $1$, and the noiseless channel induced by any unitary reaches this bound. Moreover, Eq.~\eqref{eq: block structure} implies
\begin{equation}
    \textrm{spec}(\Lambda)=\{1\}\cup\textrm{spec}(T),
\end{equation}
where eigenvalues are counted with algebraic multiplicity.

We model the implemented channel as
\begin{equation}\label{eq: noise structure}
    \tilde{\mathcal{U}}=\mathcal{U}\Lambda,
\end{equation}
where the noise channel $\Lambda$ acts immediately before the ideal unitary channel $\mathcal{U}$. The detectable spectral quantity is usually the spectrum of $\tilde{\mathcal{U}}$, rather than that of $\Lambda$ alone. This distinction is essential when interpreting the spectral information.

Throughout this article, fidelity refers to process fidelity. For a channel $\Lambda$, its process fidelity relative to the identity is~\cite{Nielsen_2002}
\begin{equation}\label{eq: fidelity def}
    F(\Lambda)=\frac{1}{d^2}\tra{\Lambda}=\frac{1}{d^2}\sum_{P\in\mathbb{P}}\Lambda_{PP}.
\end{equation}
Under Eq.~\eqref{eq: noise structure}, one can also define the fidelity of the implemented channel relative to its ideal target:
\begin{equation}\label{eq: unitary fidelity}
    F(\tilde{\mathcal{U}},\mathcal{U})=F(\Lambda)=\frac{1}{d^2}\tra{\mathcal{U}^\dagger\tilde{\mathcal{U}}}.
\end{equation}
The fidelity lies between $0$ and $1$, and the infidelity is defined as $r=1-F$.

Generally, the spectrum of $\tilde{\mathcal{U}}$ alone does not determine its fidelity. For example, suppose the ideal target $U=Z$ and the implemented unitary is
\begin{equation}
    V_\theta=\cos\theta\,Z+\sin\theta\,X.
\end{equation}
One can verify that all the corresponding unitary channels have the same spectrum $\{1,1,-1,-1\}$. However, $F(\mathcal{V}_\theta,\mathcal{U})=\cos^2\theta$, which varies with $\theta$. Thus, even exact agreement between the ideal and implemented spectrum does not imply high fidelity. Without additional assumptions, spectral information generally provides fidelity upper bounds rather than a unique fidelity value~\cite{Helsen_2019}.

\subsection{Spectral tomography and the matrix-pencil method}\label{subsec: matrix pencil}

Spectral tomography estimates eigenvalues of a channel from the signal produced by repetitions of this channel~\cite{Helsen_2019}. For an initial state $\rho$ and a measurement $M$, repeating $\tilde{\mathcal{U}}$ a total of $m$ times gives
\begin{equation}
    f(m)=\lbra{M}\tilde{\mathcal{U}}^m\lket{\rho}.
\end{equation}
In this article, we always assume a PTM is diagonalizable. The discussion about non-diagonalizable cases can be found in~\cite{Helsen_2019}. The signal then takes the form
\begin{equation}\label{eq: spectral tomography principle}
    f(m)=\sum_j a_j\lambda_j^m,\ \lambda_j\in\textrm{spec}(\tilde{\mathcal{U}}).
\end{equation}
We refer to each $\lambda_j^m$ as a mode, and $a_j$ as the corresponding amplitude. Fixed SPAM errors change the amplitudes, but not the eigenvalues of the repeated channel.

We use the matrix-pencil method~\cite{56027} to estimate the eigenvalues from the signal $\{\hat f(0),\ldots,\hat f(K)\}$, where $\hat{f}$ deviates from $f$ due to sampling fluctuations, and the maximum sequence length $K$ determines both the longest sequence used in the experiment and the number of signal values available for the fitting. We first arrange these values in a Hankel matrix,
\begin{equation}
    Y_{ij}=\hat f(i+j),\ 0\leqslant i\leqslant K-l,\ 0\leqslant j\leqslant l.
\end{equation}
Here $1\leqslant l\leqslant K$ is chosen to balance the matrix dimensions. We usually choose $l=\lceil K/2\rceil$. An exact signal with $r$ distinct visible modes has Hankel rank $\textrm{rank}(Y)=r$, provided both dimensions are sufficiently large. However, one often cannot determine in advance how many different modes to use for the fitting. We instead choose a fitting \textit{order} $r'$ (the chosen $r'$ should be no larger than $r$ when we know the value of $r$ in advance). We compute the singular value decomposition $Y=W\Sigma Z^\dagger$ and retain its largest $r'$ singular values. Define the reduced matrix $B=\Sigma_{r'}(Z^\dagger)_{1:r',:}$, where $\Sigma_{r'}$ is the retained $r'\times r'$ diagonal block, and obtain $X_0$ and $X_1$ by deleting its last and first columns. The eigenvalue estimates are
\begin{equation}
    \{\hat\lambda_{j}\}_{j=1}^{r'}=\textrm{spec}(X_1X_0^+),
\end{equation}
where $+$ denotes the Moore-Penrose pseudoinverse. The amplitudes are subsequently obtained by linear least squares with the estimated modes held fixed. It is easy to see that the complexity of the entire algorithm is a polynomial in terms of $K$.

This construction requires $l\geqslant r'$ and $K-l+1\geqslant r'$, giving the algebraic condition $K\geqslant2r'-1$. However, many factors, such as sampling fluctuations, mode separation and amplitudes can affect the effectiveness of the sequence length. An effective $K$ is usually much larger than $2r'-1$.

\subsection{KAK decomposition and group twirling}\label{subsec: kak}

Up to an irrelevant global phase, any two-qubit unitary admits the KAK decomposition~\cite{PhysRevA.67.042313,tucci2005introductioncartanskakdecomposition}
\begin{equation}
    U=LAR,
\end{equation}
where $L$ and $R$ are local unitaries, which can be decomposed as $L=L_1\otimes L_2,\ R=R_1\otimes R_2$, and
\begin{equation}\label{eq: def of A}
    A=\exp[-i(c_xXX+c_yYY+c_zZZ)]
\end{equation}
is the nonlocal part, where $c_x,c_y,c_z$ are called Cartan angles.

For a finite unitary group $\mathbb{G}$, one can apply group twirling to a channel $\Lambda$,
\begin{equation}
    \Lambda_{\mathbb{G}}=\mathbb{E}_{G\in\mathbb{G}}\mathcal{G}^\dagger\Lambda\mathcal{G}.
    \label{eq: twirl def}
\end{equation}
By cyclicity of the matrix trace, we have
\begin{equation}
    \begin{gathered}
        F(\Lambda_{\mathbb{G}})=\frac{1}{d^2}\tra{\Lambda_{\mathbb{G}}}=\frac{1}{d^2}\tra{\Lambda}=F(\Lambda).
    \end{gathered}
\end{equation}
Twirling an implemented gate $\tilde{\mathcal{U}}$ is not generally equivalent to twirling its noise channel $\Lambda$. Different twirling groups lead to different twirling effects. Next, we introduce two kinds of twirling and their related techniques used in our protocols.

\paragraph{Pauli twirling and character filtering.}
Define the rank-one Pauli projectors
\begin{equation}
    \Pi_P=\lketbra{\sigma_P}{\sigma_P}.
\end{equation}
Twirling over the full Pauli group removes all off-diagonal PTM elements,
\begin{equation}
    \Lambda_{\mathbb{P}}=\sum_{P\in\mathbb{P}}\Lambda_{PP}\Pi_P.
\end{equation}
The diagonal entries are called the Pauli eigenvalues of the twirled channel. Trace preservation fixes $\Lambda_{II}$ to $1$, leaving $d^2-1$ nontrivial parameters.

For Pauli operators $P,Q\in\mathbb{P}$, define the commutation character
\begin{equation}
    \chi_P(Q)=\begin{cases}
        +1, &PQ=QP,\\
        -1, &PQ=-QP.
    \end{cases}
\end{equation}
It follows that
\begin{equation}\label{eq: pauli character filter}
    \begin{gathered}
        \mathcal{Q}=\sum_{P\in\mathbb{P}}\chi_P(Q)\Pi_P,\\
        \Pi_P=\frac{1}{d^2}\sum_{Q\in\mathbb{P}}\chi_P(Q)\mathcal{Q}.
    \end{gathered}
\end{equation}

Equation~\eqref{eq: pauli character filter} is a simple but widely used example of character benchmarking, which provides an operational method for selecting a single Pauli component through classical weighting~\cite{helsen2019newclassefficientrandomized}. As an example, consider the signal obtained by applying a Pauli-twirled channel $\Lambda_{\mathbb{P}}$ $m$ times,
\begin{equation}
    f(m)=\lbra{M}\Lambda_{\mathbb{P}}^m\lket{\rho}=\sum_{P\in\mathbb{P}}\lbra{M}\Pi_P\lket{\rho}\Lambda_{PP}^m,
\end{equation}
which consists of $d^2$ exponential terms. When we insert a character-weighted Pauli at the beginning of the circuit, we obtain
\begin{equation}
    \begin{aligned}
        f_P(m)&=\mathbb{E}_{Q\in\mathbb{P}}\chi_P(Q)\lbra{M}\Lambda_{\mathbb{P}}^m\mathcal{Q}\lket{\rho}\\
        &=\lbra{M}\Lambda_{\mathbb{P}}^m\Pi_P\lket{\rho}\\
        &=\lbra{M}\Pi_P\lket{\rho}\Lambda_{PP}^m,
    \end{aligned}
\end{equation}
which is a single exponential term. This procedure is called character filtering, and $\Pi_P$ is called a character filter.

For a channel that is only block diagonal rather than completely diagonal, the selected filter may overlap several eigen-operators within its block. Character filtering then yields a multi-exponential signal rather than a single exponential, depending on the size of the block. Equation~\eqref{eq: multi exp example} shows an example, where $\Lambda=\Lambda_1\oplus\Lambda_2$, and $P$ belongs to the subspace of $\Lambda_1$.
\begin{equation}\label{eq: multi exp example}
    \begin{aligned}
        f'_P(m)&=\mathbb{E}_{Q\in\mathbb{P}}\chi_P(Q)\lbra{M}\Lambda^m \mathcal{Q}\lket{\rho}\\
        &=\lbra{M}\Lambda^m \Pi_P\lket{\rho}\\
        &=\lbra{M}\Lambda_1^m \Pi_P\lket{\rho}\\
        &=\sum_j a_j\lambda_j^m,\ \lambda_j\in\textrm{spec}(\Lambda_1).
    \end{aligned}
\end{equation}

\paragraph{Twirling over $\mathbb{S}$.}
In this article, we often use the following two-qubit Pauli subgroup
\begin{equation}
    \mathbb{S}=\{II,XX,YY,ZZ\},
\end{equation}
where tensor-product symbols are omitted. $\mathbb{S}$-twirling decomposes the operator space into the following four four-dimensional sectors, and we refer to these as KAK sectors.
\begin{equation}\label{eq: kak sectors}
    \begin{aligned}
        V_0&=\textrm{span}\{II,XX,YY,ZZ\},\\
        V_x&=\textrm{span}\{XI,IX,YZ,ZY\},\\
        V_y&=\textrm{span}\{YI,IY,XZ,ZX\},\\
        V_z&=\textrm{span}\{ZI,IZ,XY,YX\}.
    \end{aligned}
\end{equation}
Denoting their orthogonal projectors by $\Pi_\chi$, this twirling has the form
\begin{equation}\label{eq: S twirling effect}
    \Lambda_{\mathbb{S}}=\sum_{\chi\in\{0,x,y,z\}}\Pi_\chi\Lambda\Pi_\chi.
\end{equation}
It removes elements between different sectors while preserving the full matrix within each sector. Unlike full Pauli twirling, it does not generally diagonalize the PTM.

Both Pauli twirling and $\mathbb{S}$-twirling can be implemented using only local auxiliary gates. As mentioned in Sec.~\ref{sec: intro}, restricting auxiliary operations to local gates avoids additional multi-qubit compilation and limits the implementation overhead. Their regular implementation also makes gate-independent noise models more plausible than when different operations require substantially different compiled circuits, such as the Clifford group. These considerations motivate the use of local auxiliary gates, although locality alone does not guarantee error-free operations or gate-independent noise.

The operators $XX$, $YY$, and $ZZ$ commute pairwise, so $A$ defined in Eq.~\eqref{eq: def of A} commutes with every element of $\mathbb{S}$ and admits
\begin{equation}
    \begin{aligned}
        \mathcal{A}_{\mathbb{S}}&=\frac{1}{4}\sum_{P\in\mathbb{S}}\mathcal{P}^\dag\mathcal{A}\mathcal{P}\\
        &=\frac{1}{4}\sum_{P\in\mathbb{S}}\mathcal{A}\mathcal{P}^\dag\mathcal{P}\\
        &=\mathcal{A}.
    \end{aligned}
\end{equation}
Recall the $\mathbb{S}$-twirling effect described by Eq.~\eqref{eq: S twirling effect}, $\mathcal{A}$ is naturally block diagonal, with blocks defined in Eq.~\eqref{eq: kak sectors}. This is why we refer to these blocks as KAK sectors. Furthermore, $\mathcal{A}(P)=P$ for any $P\in\mathbb{S}$, so
\begin{equation}
    \mathcal{A}=\mathcal{A}|_{V_0}\oplus\mathcal{A}|_{V_x}\oplus\mathcal{A}|_{V_y}\oplus\mathcal{A}|_{V_z}=I_4\oplus\mathcal{A}|_{V_x}\oplus\mathcal{A}|_{V_y}\oplus\mathcal{A}|_{V_z}.
\end{equation}
The ideal eigenvalues of these blocks are given in Appendix~\ref{sec: ideal spectrum}.

\section{KAK spectral tomography}\label{sec: kst}

\subsection{Motivation}\label{subsec: sqt}
Spectral quantum tomography (SQT) reconstructs the spectrum of an implemented gate from the signal obtained by summing Pauli measurements~\cite{Helsen_2019}. With ideal SPAM, the two-qubit case is
\begin{equation}\label{eq: sqt trace}
    \begin{aligned}
        g_{\mathrm{SQT}}(m)&=\sum_{P\neq II}\lbra{\sigma_P}\tilde{\mathcal{U}}^m\lket{\sigma_P}\\
        &=\tra{T^m}\\
        &=\sum_{j=1}^{15}\lambda_j^m,\ \lambda_j\in\textrm{spec}(T),
    \end{aligned}
\end{equation}
where $T$ is the block shown in Eq.~\eqref{eq: block structure}. Here the initial state $\lket{\sigma_P}$ can be realized as $\Pi_P\lket{\rho}$, with character filter $\Pi_P$ and some initial state $\rho$. $\lbra{\sigma_P}$ can be realized as measuring in the $P$ basis. For small SPAM errors, the mode amplitudes are expected to remain close to their ideal values of $1$ (here we temporarily treat the identical eigenvalues separately), ensuring the visibility of each mode.

One of the major challenges that SQT faces is the difficulty in handling degenerate spectrum. When the ideal $\mathcal{U}$ has multiple identical eigenvalues, small noise would split them into several nearly indistinguishable terms. Separating them then becomes sensitive to sampling fluctuations.

This observation motivates KAK spectral tomography (KST). Rather than obtaining the entire spectrum from one signal, we use the KAK structure to construct four independent signals, each of which contains at most three or four modes. This method significantly reduces the near degeneracy, although it cannot completely eliminate it.

\subsection{Protocol}\label{subsec: kst protocol}
We assume that all local auxiliary gates are ideal. Given $U=LAR$, the implemented channel is modeled as
\begin{equation}\label{eq: kak noise structure}
    \begin{gathered}
        \tilde{\mathcal{U}}=\mathcal{LA}\Lambda\mathcal{R},\\
        \tilde{\mathcal{A}}=\mathcal{A}\Lambda=\mathcal{L}^\dag\tilde{\mathcal{U}}\mathcal{R}^\dag.
    \end{gathered}
\end{equation}
This definition differs slightly from Eq.~\eqref{eq: noise structure}. One can absorb the conjugation $\mathcal R\Lambda\mathcal R^\dag$ into $\Lambda$ to make them consistent, while maintaining its fidelity unchanged.

Each application of $\tilde{\mathcal A}$ requires one application of the target gate, preceded by $R^\dagger$ and followed by $L^\dagger$. Then we apply $\mathbb{S}$-twirling to $\tilde{\mathcal{A}}$ to obtain four blocks, as explained in Eq.~\eqref{eq: S twirling effect}. For each block, we use character filtering to isolate it and implement standard SQT. Specifically, when handling $V_x$, we estimate
\begin{equation}
    \begin{aligned}
        \textrm{tr}[(\tilde{\mathcal{A}}_{\mathbb{S}}|_{V_x})^m]&=\textrm{tr}(\tilde{\mathcal{A}}_{\mathbb{S}}^m|_{V_x})\\
        &=\sum_{P\in\{IX,XI,YZ,ZY\}}\lbra{\sigma_{P}}\tilde{\mathcal{A}}_{\mathbb{S}}^m\lket{\sigma_{P}},
    \end{aligned}
\end{equation}
and implement the matrix-pencil method to evaluate the eigenvalues of $\tilde{\mathcal{A}}_\mathbb{S}|_{V_x}$. The same goes for $V_y$ and $V_z$. When handling $V_0$, we omit the $II$ term as SQT does, and estimate
\begin{equation}
    \sum_{P\in\{XX,YY,ZZ\}}\lbra{\sigma_{P}}\tilde{\mathcal{A}}_{\mathbb{S}}^m\lket{\sigma_{P}}.
\end{equation}
A more formal and concrete description is shown in Algorithm~\ref{algo: kst}.

\begin{algorithm}[h]
    \caption{KAK spectral tomography}
    1. Given target gate $U$, calculate the KAK decomposition $U=LAR$. Choose the maximum sequence length $K$.

    2. For each $P\neq II$ and $0\leqslant m\leqslant K$, perform steps 3--5.

    3. Suppose $P=P^{(1)}\otimes P^{(2)}$. Let $\ket{\psi_i}$ be the $+1$ eigenstate of $P^{(i)}$, and prepare the initial state $\ket{\psi_1}\ket{\psi_2}$. Because of preparation error, the actual state $\rho$ may slightly deviate from $\ketbra{\psi_1\psi_2}$.

    4. Uniformly sample $Q\in\mathbb{P}$ and $P_1,P_2,\cdots,P_m\in\mathbb{S}$. Implement
    \begin{equation}
        \prod_{i=1}^m(\mathcal{P}_i^\dag\tilde{\mathcal{A}}\mathcal{P}_i)\mathcal{Q}=\prod_{i=1}^m(\mathcal{P}_i^\dag\mathcal{L}^\dag\tilde{\mathcal{U}}\mathcal{R}^\dag\mathcal{P}_i)\mathcal{Q}.
    \end{equation}
    Then measure the final state by observable $P$, that is, measure the $i$-th qubit in the $P^{(i)}$ basis. Because of measurement error, the actual observable $M$ may slightly deviate from $P$. Repeat the circuit several times to estimate
    \begin{equation}
        g_P(m,Q,\{P_i\})=\lbra{M}\prod_{i=1}^m(\mathcal{P}_i^\dag\tilde{\mathcal{A}}\mathcal{P}_i)\mathcal{Q}\lket{\rho}.
    \end{equation}

    5. Repeat step 4 for random $Q$ and $\{P_i\}$, and obtain an estimate of
    \begin{equation}
        \begin{aligned}
            g_P(m)&=\mathbb{E}\chi_P(Q)g_P(m,Q,\{P_i\})\\
            &=\lbra{M}\tilde{\mathcal{A}}_{\mathbb{S}}^m\Pi_P\lket{\rho}.
        \end{aligned}
    \end{equation}
    Without SPAM error, $g_P(m)$ should just be $\lbra{\sigma_P}\tilde{\mathcal{A}}_{\mathbb{S}}^m\lket{\sigma_P}$. When SPAM error exists, because $\tilde{\mathcal{A}}_{\mathbb{S}}$ is block diagonal, $g_P(m)$ only involves the modes within the KAK sector to which $P$ belongs.

    6. For each $\chi\in\{0,x,y,z\}$ and $0\leqslant m\leqslant K$, calculate
    \begin{equation}
        g_\chi(m)=\sum_{\substack{P\in V_\chi\\P\neq II}}g_P(m),
    \end{equation}
    and implement the matrix-pencil algorithm to each sector signal.
    \label{algo: kst}
\end{algorithm}

There are two points that need to be noted. The first is that KST protocol estimates the eigenvalues of $\tilde{\mathcal{A}}_\mathbb{S}$, rather than $\tilde{\mathcal{A}}$ as standard SQT protocol does (neglecting $L$ and $R$). So one should be careful when comparing the results of these two protocols.

The second is that because $\mathcal{A}$ is naturally block diagonal, under the influence of small noise, $\mathcal{A}$, $\tilde{\mathcal{A}}$ and $\tilde{\mathcal{A}}_\mathbb{S}$ should be close to each other, and the moduli of their eigenvalues should all be close to $1$, because that of $\mathcal{A}$ are strictly $1$'s. This ensure for quite large values of $m$, the spectral signals remain visible. Stronger twirling, such as Pauli twirling, can destroy this property and cause the spectral signals to decay more quickly.

\subsection{Spectrum reconstruction}\label{subsec: mode selection}

\begin{figure}[h!]
    \centering
    \subfigure[Modes of two models]{%
        \begin{minipage}[b]{0.40\linewidth}
            \centering
            \input{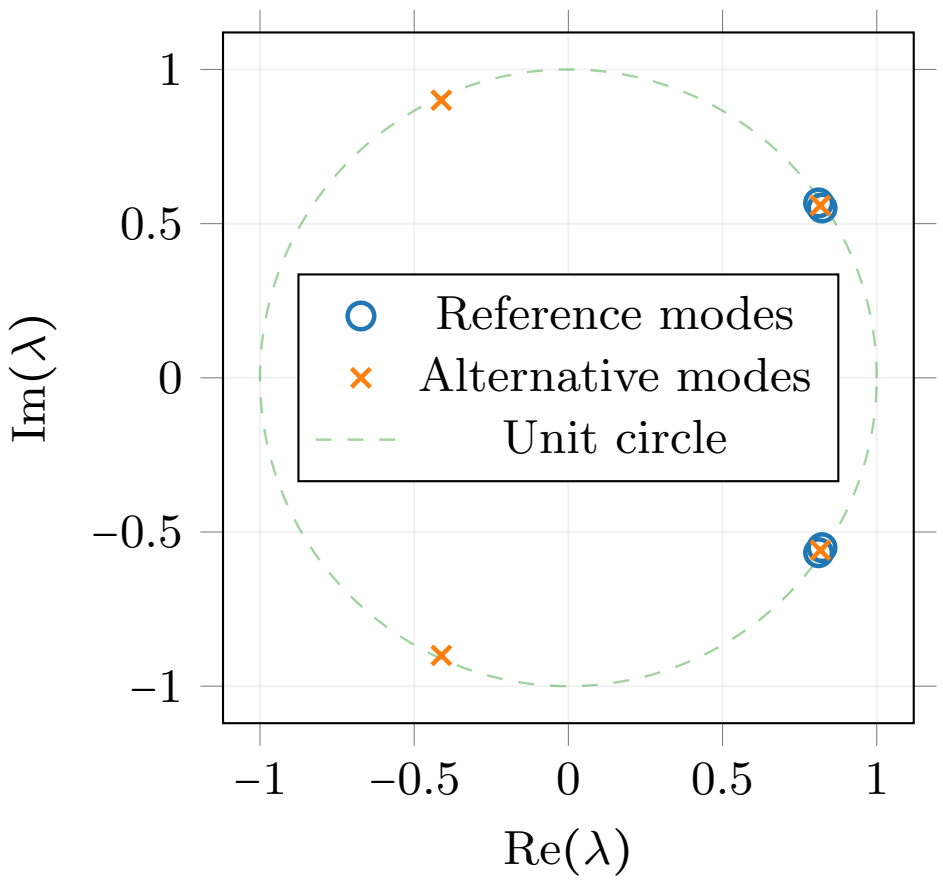}
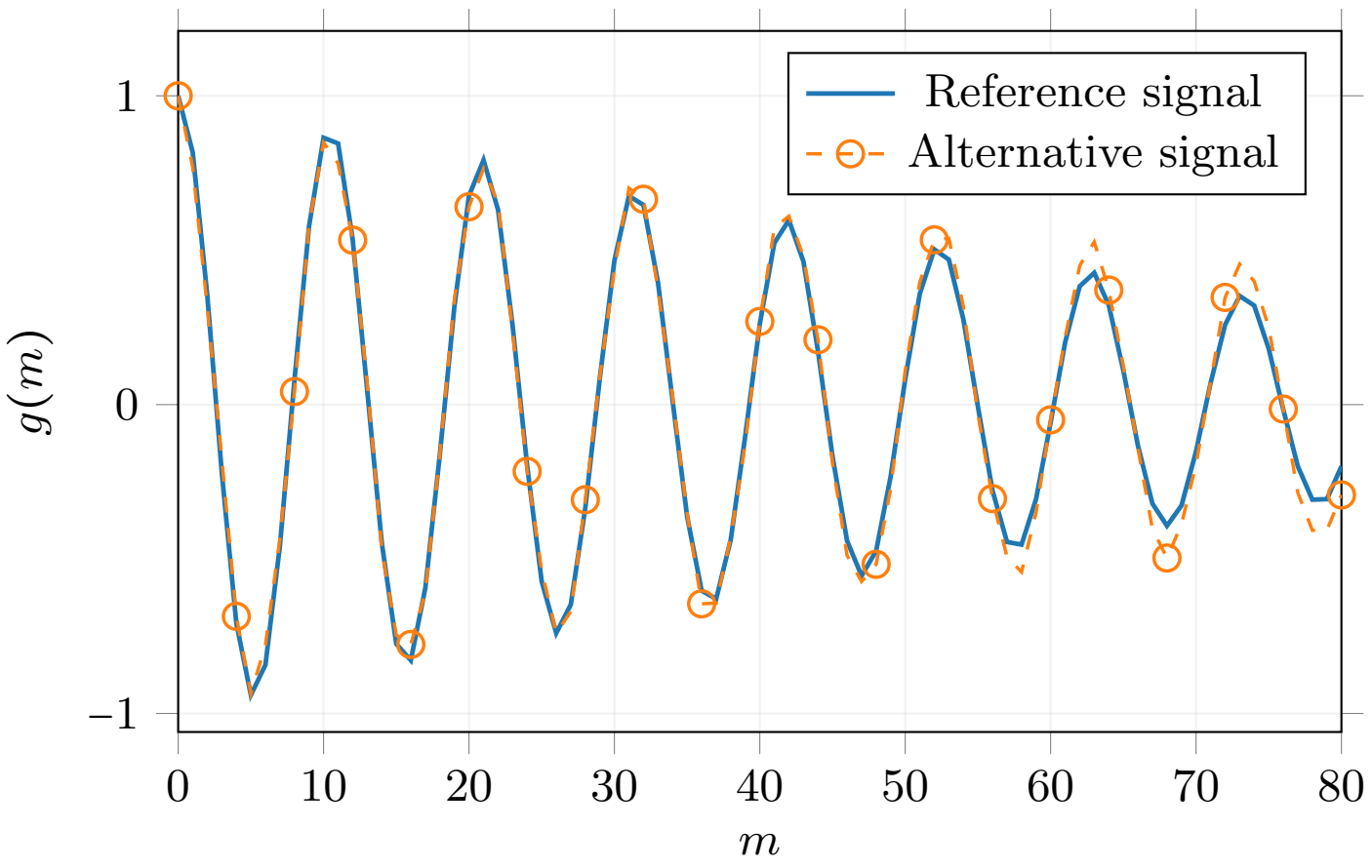
        \end{minipage}%
    }
    \hfill
    \subfigure[Signals of two models]{%
        \begin{minipage}[b]{0.585\linewidth}
            \centering
\begingroup
\small
\definecolor{nearReference}{HTML}{1F77B4}
\definecolor{nearAlternative}{HTML}{FF7F0E}
\definecolor{nearCircle}{HTML}{2CA02C}
\pgfmathsetlengthmacro{\nearAxisWidth}{\linewidth-8em}
\pgfmathsetlengthmacro{\nearSignalHeight}{0.603*(\linewidth-8em)}
\begin{tikzpicture}[font=\small]
\begin{axis}[
  scale only axis,width=\nearAxisWidth,
  font=\small,label style={font=\small},
  tick label style={font=\small},
  legend style={font=\small},
  grid=major,grid style={opacity=0.2},tick align=outside,
  enlargelimits=false,
  height=\nearSignalHeight,
  xmin=0,xmax=80,ymin=-1.06,ymax=1.21,
  xlabel={$m$},ylabel={$g(m)$},legend pos=north east,
]
\addplot[color=nearReference,line width=0.9pt,no marks] coordinates {
(0,1)
(1,0.8170414049880933)
(2,0.3550758081631693)
(3,-0.22035476843433038)
(4,-0.70777086653120336)
(5,-0.94029641763033855)
(6,-0.84276099919748859)
(7,-0.45583604002757649)
(8,0.080480994989375915)
(9,0.57745627193627547)
(10,0.86402260918578633)
(11,0.84563756152133918)
(12,0.53535557813726165)
(13,0.046947496696458833)
(14,-0.44671598778117172)
(15,-0.77482592767076008)
(16,-0.82771122422535681)
(17,-0.59340616870049523)
(18,-0.15955089700519443)
(19,0.31923956780020268)
(20,0.67643607864990396)
(21,0.79149111390468574)
(22,0.63045047258406173)
(23,0.25562580320100314)
(24,-0.19828195743421387)
(25,-0.57253178891963907)
(26,-0.73982918887253146)
(27,-0.64756527583317214)
(28,-0.33414265249040043)
(29,0.086592362151906493)
(30,0.46662122601480893)
(31,0.67579325000311807)
(32,0.6463498520854507)
(33,0.39471953036247237)
(34,0.013632569410693601)
(35,-0.36193925941241711)
(36,-0.60254381808490598)
(37,-0.62882416893351434)
(38,-0.43757547357540483)
(39,-0.10077177986490986)
(40,0.26136374603273471)
(41,0.52321899126344351)
(42,0.59732156719078833)
(43,0.46346685141161859)
(44,0.17378041318191345)
(45,-0.16735211900001723)
(46,-0.44083071879959845)
(47,-0.55438031333814142)
(48,-0.47361069658430505)
(49,-0.23217005631942042)
(50,0.081898692752023605)
(51,0.35817518658710074)
(52,0.50263805537052986)
(53,0.46959897448772503)
(54,0.27597122398241991)
(55,-0.0065122980612019044)
(56,-0.27775923919224993)
(57,-0.44473272797434937)
(58,-0.45330773815101977)
(59,-0.30568110488854072)
(60,-0.057786849255537578)
(61,0.20174399404748275)
(62,0.3832128832620379)
(63,0.42680493286413668)
(64,0.32219984464924889)
(65,0.11045274437698141)
(66,-0.13190606424705725)
(67,-0.32045979741763192)
(68,-0.39226030928086941)
(69,-0.32675875181862701)
(70,-0.15138383521711565)
(71,0.069616121738471065)
(72,0.25862305011579101)
(73,0.35186049974869349)
(74,0.32084378946399644)
(75,0.18088044708978904)
(76,-0.015833922791565147)
(77,-0.19957061932543407)
(78,-0.30773183404493576)
(79,-0.30611756855915506)
(80,-0.19959215634863972)
};
\addlegendentry{Reference signal}
\addplot[color=nearAlternative,dashed,line width=0.65pt,mark=o,mark size=2.5pt,mark options={solid,fill=none},mark repeat=4,mark phase=1] coordinates {
(0,1)
(1,0.76791955368289133)
(2,0.31531551724374718)
(3,-0.17436971590880074)
(4,-0.68559463258733222)
(5,-0.93573204020521905)
(6,-0.77873016782979798)
(7,-0.43357899717333365)
(8,0.042162645645642156)
(9,0.58073920153875103)
(10,0.84838884842801676)
(11,0.78093627919767816)
(12,0.53270350001453393)
(13,0.068160547435641997)
(14,-0.46653504275528668)
(15,-0.74697357621675042)
(16,-0.77647108331662695)
(17,-0.60661381634799427)
(18,-0.1599556403487542)
(19,0.3436482903673766)
(20,0.64076651617662317)
(21,0.76394406463353914)
(22,0.65224058201915525)
(23,0.23831853228145608)
(24,-0.21608540126358827)
(25,-0.537242578172761)
(26,-0.73995767668946399)
(27,-0.6703203131482528)
(28,-0.30801802765510927)
(29,0.090312978396619684)
(30,0.44071904552764168)
(31,0.7009619510510301)
(32,0.66504318264985174)
(33,0.37184401702113717)
(34,0.026391744914595536)
(35,-0.35206665282063626)
(36,-0.645006033006992)
(37,-0.64275285073928246)
(38,-0.42983427039702199)
(39,-0.12778074799216749)
(40,0.26945938748172216)
(41,0.57283981423055674)
(42,0.61017514535677808)
(43,0.47952599881449742)
(44,0.21008060664653444)
(45,-0.18982260848193844)
(46,-0.48807293596864298)
(47,-0.57275209227043822)
(48,-0.51704685511146509)
(49,-0.27263401750049543)
(50,0.11045846650652953)
(51,0.39641873939195837)
(52,0.53357044926607988)
(53,0.53861776844402875)
(54,0.31761321054678604)
(55,-0.030324940806524654)
(56,-0.30433205268364055)
(57,-0.49314429717812408)
(58,-0.54195313708165205)
(59,-0.34897746835146798)
(60,-0.049396154408615313)
(61,0.217508925144937)
(62,0.45002377369621666)
(63,0.52712436278659169)
(64,0.37106223637726704)
(65,0.12555065535962415)
(66,-0.13971605730305461)
(67,-0.40195087471030844)
(68,-0.49665965601073253)
(69,-0.3872627339155203)
(70,-0.19392650139589826)
(71,0.072273059730511818)
(72,0.34714235858474352)
(73,0.45491051439960994)
(74,0.39920066175133956)
(75,0.2504834535959452)
(76,-0.014280168020969316)
(77,-0.28528204811422986)
(78,-0.40693899075889767)
(79,-0.40657578621411533)
(80,-0.29248494460926139)
};
\addlegendentry{Alternative signal}
\end{axis}
\end{tikzpicture}
\endgroup
        \end{minipage}%
    }
    \caption{Demonstration of the ambiguity of multi-exponential reconstruction near degeneracy. The reference model contains two nearby complex-conjugate mode pairs with equal amplitudes. The alternative model assigns dominant amplitudes to a pair near the reference modes, and assigns small amplitudes to a distant pair. Although the mode sets differ substantially, their signals nearly coincide before $m$ reaches $50$.}
    \label{fig: near degeneracy}
\end{figure}

When applying the matrix-pencil method, the choice of order requires particular care in the near-degenerate case. As shown in Fig.~\ref{fig: near degeneracy}, a full-order fit (the order equals the known upper bound of the number of modes) can assign small amplitudes to some obviously incorrect modes while still reproducing the signal. We therefore retain the full-order fit when it is consistent with the signal and the known ideal spectrum, and consider a lower-order fitting when the full-order result fails these checks. Specifically, we check whether the root-mean-square fit residual is within a tolerance and each ideal eigenvalue can be matched to an estimated eigenvalue within a prescribed distance.

When a lower-order fit is adopted, to ensure that the number of estimated eigenvalues is consistent with the ideal spectrum, we should introduce multiplicity, the number of times an estimated eigenvalue is included in the expanded spectrum. Equation~\eqref{eq: sqt trace} suggests that the amplitude of each mode should be near to its multiplicity when errors are small, so we assign multiplicities from the magnitudes of the fitted amplitudes, requiring equal multiplicities for complex-conjugate pairs. Finally, we match the expanded list one-to-one to the known ideal spectrum. The precise rule is given in Appendix~\ref{subsec: multiplicity}.

\section{Sector echo benchmarking and hybrid estimation}\label{sec:seb}

Many randomized benchmarking protocols construct a circuit unit that is repeated to probe noise accumulation. For example, equivalently, standard RB constructs $\mathcal{C}^\dag\Lambda \mathcal{C}$ and cycle benchmarking constructs $\mathcal{U}\mathcal{P}^\dag\Lambda\mathcal{P}$~\cite{PhysRevLett.106.180504,Erhard_2019}, where $\Lambda$ is the noise channel, $U$ is the target gate, $C$ and $P$ are random Clifford and random Pauli gates respectively. Sector echo benchmarking (SEB) follows this approach but focuses on a selected sector of the PTM of the unit. The noiseless unit acts as the identity within this sector, while its action elsewhere may remain nontrivial. This structure motivates the name ``sector echo".
\subsection{Protocol}\label{subsec: seb protocol}
We retain the noise model described by Eq.~\eqref{eq: kak noise structure}. Local auxiliary gates are assumed ideal. Consider $V_x$. Define $J_x=IX$, $J_y=IY$, and $J_z=IZ$. Because $IX$ commutes with $XX$ and anti-commutes with $YY,ZZ$,
\begin{equation}
    \begin{gathered}
        J_x^\dag AJ_x=\exp[-i(c_xXX-c_yYY-c_zZZ)],\\
        J_x^\dag AJ_xA=\exp(-2ic_xXX).
    \end{gathered}
\end{equation}
Although $e^{-2ic_xXX}$ is generally not the identity, every operator in $V_x$ commutes with $XX$. Its conjugation action is therefore the identity on $V_x$. The same construction applies to the other two sectors.
\begin{equation}\label{eq: echo}
    (\mathcal{J}_\chi^\dag\mathcal{A}\mathcal{J}_\chi\mathcal{A})|_{V_\chi}=I_4,\ \chi\in\{x,y,z\}.
\end{equation}
In particular, since $\mathcal A|_{V_0}=I_4$ already, no other operation is needed when constructing an echo on $V_0$.

For the noisy implementation, we first apply independent $\mathbb S$-twirling to each occurrence of $\tilde{\mathcal A}$, then apply Pauli twirling to the entire echo. The resulting average channels are
\begin{equation}
    \begin{gathered}
        \mathcal{C}_0=\tilde{\mathcal{A}}_\mathbb{P},\\
        \mathcal{C}_\chi=(\mathcal{J}_\chi^\dag\tilde{\mathcal{A}}_\mathbb{S}\mathcal{J}_\chi\tilde{\mathcal{A}}_\mathbb{S})_\mathbb{P},\ \chi\in\{x,y,z\}.
    \end{gathered}
\end{equation}
Each $\mathcal C_\chi$ is diagonal through Pauli twirling. Then one can estimate the following signals by inserting a Pauli filter at the beginning of the circuit.
\begin{equation}
    \begin{aligned}
        g_{\chi,P}(m)&=\lbra{M}\mathcal{C}_\chi^m\Pi_P\lket{\rho}\\
        &=\lbra{M}\Pi_P\lket{\rho}(\mathcal{C}_\chi)_{PP}^m,
    \end{aligned}
\end{equation}
for $\chi\in\{0,x,y,z\}$ and $P\in V_\chi$. A simple exponential fitting can extract $(\mathcal{C}_\chi)_{PP}$ from $g_{\chi,P}(m)$. A more formal and concrete description is shown in Algorithm~\ref{algo: seb}. It should be noted that this algorithm stops at obtaining the value of $(\mathcal{C}_\chi)_{PP}$. Their relationship with infidelity is explained in Sec.~\ref{subsec: echo infidelity} and~\ref{subsec: hybrid infidelity}.

\begin{algorithm}[h]
    \caption{Sector echo benchmarking}
    1. Given target gate $U$, calculate the KAK decomposition $U=LAR$. Choose the maximum sequence length $K$.

    2. For each $\chi\in\{0,x,y,z\}$, nonidentity $P\in V_\chi$, and $0\leqslant m\leqslant K$, perform steps 3--5.

    3. Suppose $P=P^{(1)}\otimes P^{(2)}$. Let $\ket{\psi_i}$ be the $+1$ eigenstate of $P^{(i)}$, and prepare the initial state $\ket{\psi_1}\ket{\psi_2}$. Because of preparation error, the actual state $\rho$ may slightly deviate from $\ketbra{\psi_1\psi_2}$.

    4. Uniformly sample $Q\in\mathbb P$ and $P_1,P_2,\cdots,P_m\in\mathbb{P}$. If $\chi\neq 0$, also sample $H_{i,1}, H_{i,2}\in\mathbb S$ for $1\leqslant i\leqslant m$. Implement
    \begin{equation}
        \mathcal{C}=\begin{cases}
            \displaystyle\prod_{i=1}^m(\mathcal{P}_i^\dag\tilde{\mathcal{A}}\mathcal{P}_i)\mathcal Q&\chi=0\\
            \displaystyle\prod_{i=1}^m[\mathcal{P}_i^\dag\mathcal{J}_\chi^\dag(\mathcal{H}_{i,1}^\dag\tilde{\mathcal{A}}\mathcal{H}_{i,1})\mathcal{J}_\chi(\mathcal{H}_{i,2}^\dag\tilde{\mathcal{A}}\mathcal{H}_{i,2})\mathcal{P}_i]\mathcal Q&\chi\neq 0
        \end{cases}.
    \end{equation}
    Then measure the final state by observable $P$, that is, measure the $i$-th qubit in the $P^{(i)}$ basis. Because of measurement error, the actual observable $M$ may slightly deviate from $P$. Repeat the circuit several times to estimate
    \begin{equation}
        g_{\chi,P}(m,Q,\{P_i\},\{H_{i,1},H_{i,2}\})=\lbra{M}\mathcal{C}\lket{\rho}.
    \end{equation}

    5. Repeat step 4 for random $Q$, $\{P_i\}$ and $\{H_{i,1},H_{i,2}\}$, and obtain an estimate of
    \begin{equation}
        \begin{aligned}
            g_{\chi,P}(m)&=\mathbb{E}\chi_P(Q)g_{\chi,P}(m,Q,\{P_i\},\{H_{i,1},H_{i,2}\})\\
            &=\lbra{M}\mathcal{C}_\chi^m\Pi_P\lket{\rho}\\
            &=\lbra{M}\Pi_P\lket{\rho}(\mathcal{C}_\chi)_{PP}^m,
        \end{aligned}
    \end{equation}
    where $\lbra{M}\Pi_P\lket{\rho}=1$ without SPAM errors. Therefore, slight SPAM errors still allow the signal $g_{\chi,P}(m)$ to be visible.

    6. Implement an exponential fitting to extract $(\mathcal{C}_{\chi})_{PP}$ from $g_{\chi,P}(m)$.
    \label{algo: seb}
\end{algorithm}

\subsection{Infidelity estimation}\label{subsec: echo infidelity}

Like most RB protocols, SEB aims to estimate the fidelity or infidelity of the target gate. We define
\begin{equation}\label{eq: def r echo}
    r_{\mathrm{echo}}=\frac{1}{16}\sum_{P\in V_0}(1-(\mathcal{C}_0)_{PP})+\frac{1}{32}\sum_{\chi=x,y,z}\sum_{P\in V_\chi}(1-(\mathcal{C}_\chi)_{PP}),
\end{equation}
where the factor $\frac{1}{16}$ matches the definition in Eq.~\eqref{eq: fidelity def}, and the factor $\frac{1}{32}$ accounts for the two applications of the target gate in the echoes on $V_x,V_y,V_z$.

In Appendix~\ref{subsec: r echo proof}, we establish an approximate relation between $r_{\mathrm{echo}}$ and the desired infidelity $r=1-F(\Lambda)$ by proving that their difference is quadratic in the PTM deviations $\Delta_\chi=I_4-\Lambda_{\mathbb{S}}|_{V_\chi}$. However, the significance of this bias depends on how it compares with $r$ itself. Consider the Cartan angle error
\begin{equation}\label{eq: Cartan error}
    \tilde{A}=\exp[-i((c_x+\delta_x)XX+(c_y+\delta_y)YY+(c_z+\delta_z)ZZ)].
\end{equation}
For this error, $\mathcal{C}_\chi|_{V_\chi}=I_4$ for every $\chi\in\{0,x,y,z\}$, giving $r_{\mathrm{echo}}=0$ while $r$ varies with $\delta_x,\delta_y,\delta_z$. From another perspective, a Cartan angle error of magnitude $\delta$ induces an $O(\delta)$ PTM deviation but an infidelity of only $O(\delta^2)$, so the quadratic bias $|r_{\mathrm{echo}}-r|$ can be of the same order as $r$ itself. This example motivates the use of complementary phase information from KST in Sec.~\ref{subsec: hybrid infidelity}.

\subsection{Hybrid estimation with KST}\label{subsec: hybrid infidelity}
We use the Cartan phase information to correct the contribution missed by the echo estimate.
\begin{equation}\label{eq: hybrid}
    r_\textrm{hybrid}=r_\textrm{echo}+\hat{\delta}_x^2+\hat{\delta}_y^2+\hat{\delta}_z^2.
\end{equation}
When $\Lambda$ only consists of Cartan angle errors, the definition of $\hat{\delta}_\chi$ is consistent with $\delta_\chi$ in Eq.~\eqref{eq: Cartan error}. For general cases, $\hat{\delta}_\chi$ are fitted from the results of the KST protocol, and we add some checks to ensure the reliability of this fitting, which is described in Appendix~\ref{subsec: delta estimation}.

Any noise channel sufficiently close to the identity can be factored into a unitary channel and a decoherent channel~\cite{CarignanDugas2019polardecomposition}. Expanding this product to first order in the coherent and dissipative error strengths gives
\begin{equation}
    \begin{gathered}
        \Lambda=I_{16}+\mathcal{L}_H+\mathcal{D}+\mathcal{R},\\
        \mathcal{L}_H(O)=-i[H,O],
    \end{gathered}
\end{equation}
where $H$ is the traceless Hermitian generator of the coherent error, $\mathcal{D}$ is the leading dissipative contribution, and $\mathcal{R}$ collects the higher-order terms. Denoting the coherent and dissipative error strengths by $\epsilon$ and $\kappa$, respectively, we have
\begin{equation}
    \begin{gathered}
        ||H||_F=O(\epsilon)\ (\textrm{or\ }||\mathcal{L}_H||=O(\epsilon)\textrm{\ equivalently}),\\
        ||\mathcal{D}||_F=O(\kappa),\\
        ||\mathcal{R}||_F=O(\epsilon^2+\epsilon\kappa+\kappa^2).
    \end{gathered}
\end{equation}
Under this model, we prove that
\begin{equation}\label{eq: hybrid accuracy}
    r=r_\textrm{echo}+\delta_x^2+\delta_y^2+\delta_z^2+O(\epsilon^3+\epsilon\kappa+\kappa^2).
\end{equation}
In other words, $r_\textrm{echo}+\delta_x^2+\delta_y^2+\delta_z^2$ includes the precise $\epsilon,\epsilon^2$ and $\kappa$ terms, which motivates Eq.~\eqref{eq: hybrid}. Here $\delta_x,\delta_y,\delta_z$ are the Pauli coefficients of $XX,YY,ZZ$ in $H$ respectively. The full proof is given in Appendix~\ref{subsec: r hybrid proof}.

\section{Numerical simulation}\label{sec: simulation}

We examine three aspects of our methods. First, we compare KST and the standard SQT for a noiseless target gate. Second, we evaluate KST over an ensemble of noisy two-qubit gates. Finally, we implement SEB over the same ensemble and combine the phase information from KST to estimate the infidelities.

\subsection{Noise model and sampling budget}\label{subsec: noise model}

For the studies over noisy gates, each setting has an independently sampled noise channel composed of coherent error, amplitude damping, and stochastic Pauli errors. By adjusting the amplitudes of these errors, we construct a noise channel ensemble whose infidelities range from $10^{-4}$ to $10^{-1}$. Both preparation and measurement errors are fixed throughout the simulations. The preparation error acts after the ideal preparation and the measurement error acts before the ideal measurement. More detailed explanations are given in Appendix~\ref{subsec: error construction}.

For every depth $m$ and nonidentity Pauli $P$, all sixteen $Q\in\mathbb P$ receive equal resources (recall step 4 of Algorithm~\ref{algo: kst} and step 4 of Algorithm~\ref{algo: seb}). For each $(P,Q)$, we independently sample $N_{\mathrm{seq}}=4$ internal random strings and collect $n_{\mathrm{shot}}=32$ shots per fixed circuit. The total allocation per depth is therefore
\begin{equation}
    N_m=15\times16\times N_{\mathrm{seq}}\times n_{\mathrm{shot}}=30720,
\end{equation}
for each protocol.

\subsection{KST versus standard SQT}\label{subsec: kst vs sqt}
The first experiment compares the performance of KST and standard SQT. We take $L=R=I_4$, $\Lambda=I_{16}$, and
\begin{equation}
    \begin{gathered}
        A=\exp[-i(c_xXX+c_yYY+c_zZZ)],\\
        (c_x,c_y,c_z)=(0.70,0.33,0.10).
    \end{gathered}
\end{equation}
Although the target gate is noiseless, SPAM errors and finite-shot fluctuations still exist. The latter is particularly likely to affect the accuracy of the fit.

For quantitative evaluation, let $\{\lambda_j\}$ denote the reference eigenvalues (the eigenvalues of the noisy target PTM $\tilde{\mathcal{A}}$ for standard SQT, or the eigenvalues of the twirled noisy target PTM $\tilde{\mathcal{A}}_\mathbb{S}$ for KST) and $\{\hat{\lambda}_j\}$ the estimated eigenvalues, both listed with algebraic multiplicity. In this noiseless experiment, the reference spectrum is just the ideal spectrum, which is the same for two methods. We define the root-mean-square error (RMSE)
\begin{equation}
    \mathrm{RMSE}=\min_{\pi}\left(\frac{1}{15}\sum_{j=1}^{15}|\hat{\lambda}_j-\lambda_{\pi(j)}|^2\right)^{1/2},
\end{equation}
where matching $\pi$ is performed separately within each sector for KST and over the entire traceless operator subspace for standard SQT.

\begin{figure}[h!]
    \centering
\begingroup
\small
\pgfmathsetlengthmacro{\kstPendingOneWidth}{\linewidth-25em}
\definecolor{kstPendingOneBlue}{RGB}{31,119,180}
\definecolor{kstPendingOneOrange}{RGB}{255,127,14}
\begin{tikzpicture}[font=\small]
\begin{axis}[
  scale only axis,width=\kstPendingOneWidth,height=0.35\linewidth,
  font=\small,label style={font=\small},
  tick label style={font=\small},
  title style={font=\small,align=center,text width=\kstPendingOneWidth},
  legend style={font=\small},legend pos=north east,
  xlabel={Maximum sequence length $K$},ylabel={RMSE},
  ymode=log,unbounded coords=jump,grid=major,grid style={opacity=0.25},
  tick align=outside,scaled x ticks=false,
  xmin=2,xmax=134,ymin=5.4154298791782796e-05,ymax=1.0011103827882997,
]
\addplot[color=kstPendingOneBlue,mark=*,mark size=2pt,line width=0.8pt,
  error bars/y dir=both,error bars/y explicit,
  error bars/error bar style={line width=0.65pt},
  error bars/error mark options={rotate=90,mark size=3pt,line width=0.65pt}]
table[x=K,y=rmse,y error minus=err_minus,y error plus=err_plus] {
K rmse q25 q75 err_minus err_plus n_finite n_trials
8 0.3721528066516494 0.24454757195641585 0.64052234130392194 0.12760523469523355 0.26836953465227253 100 100
12 0.018272424212433788 0.01214252773768806 0.030255845173595468 0.0061298964747457281 0.01198342096116168 100 100
16 0.0043782330067741182 0.0031085927747129699 0.0054090248946370937 0.0012696402320611483 0.0010307918878629755 100 100
24 0.0013115754623010791 0.0010753829969692001 0.0014703089644061108 0.00023619246533187902 0.00015873350210503169 100 100
32 0.00076019307302606354 0.00064849572768183863 0.00085698183531319983 0.00011169734534422491 9.6788762287136295e-05 100 100
48 0.00041835036687559366 0.00035779249352853174 0.00049026269785375784 6.0557873347061916e-05 7.1912330978164186e-05 100 100
64 0.00025975949419175255 0.00022513515323101756 0.00029147553416464044 3.4624340960734987e-05 3.1716039972887894e-05 100 100
96 0.00014815991420857966 0.00012478859162052831 0.0001701903450921426 2.3371322588051354e-05 2.2030430883562938e-05 100 100
128 9.5940243404147904e-05 8.4640967686948153e-05 0.00010957341863683963 1.1299275717199751e-05 1.363317523269173e-05 100 100
};
\addlegendentry{KST}
\addplot[color=kstPendingOneOrange,mark=*,mark size=2pt,line width=0.8pt,
  error bars/y dir=both,error bars/y explicit,
  error bars/error bar style={line width=0.65pt},
  error bars/error mark options={rotate=90,mark size=3pt,line width=0.65pt}]
table[x=K,y=rmse,y error minus=err_minus,y error plus=err_plus] {
K rmse q25 q75 err_minus err_plus n_finite n_trials
8 nan nan nan nan nan 0 100
12 nan nan nan nan nan 0 100
16 nan nan nan nan nan 0 100
24 nan nan nan nan nan 0 100
32 0.34310730894574526 0.33383241604637648 0.34986369997386113 0.0092748928993687807 0.0067563910281158712 100 100
48 0.0017258398282793174 0.0013282557050493624 0.0021099974131048834 0.00039758412322995497 0.000384157584825566 100 100
64 0.00068404709985228108 0.00060446738014513903 0.00092994358023699877 7.9579719707142059e-05 0.00024589648038471769 100 100
96 0.00029464300803281923 0.00024864602102527992 0.00033051842614086567 4.5996987007539309e-05 3.587541810804644e-05 100 100
128 0.00018218183212575937 0.00016104628655874294 0.00022001070648317597 2.1135545567016429e-05 3.7828874357416606e-05 100 100
};
\addlegendentry{Standard SQT}
\end{axis}
\end{tikzpicture}
\par\endgroup
    \caption{RMSE of KST and SQT computed from the same $100$-trial dataset. Points show the median across trials, and error bars span the $25$th--$75$th percentiles. SQT points at $K=8,12,16,24$ are omitted because the available data do not support a full-order fitting, and no lower-order fit passes the acceptance checks described in Sec.~\ref{subsec: mode selection}.}
    \label{fig: pending1}
\end{figure}
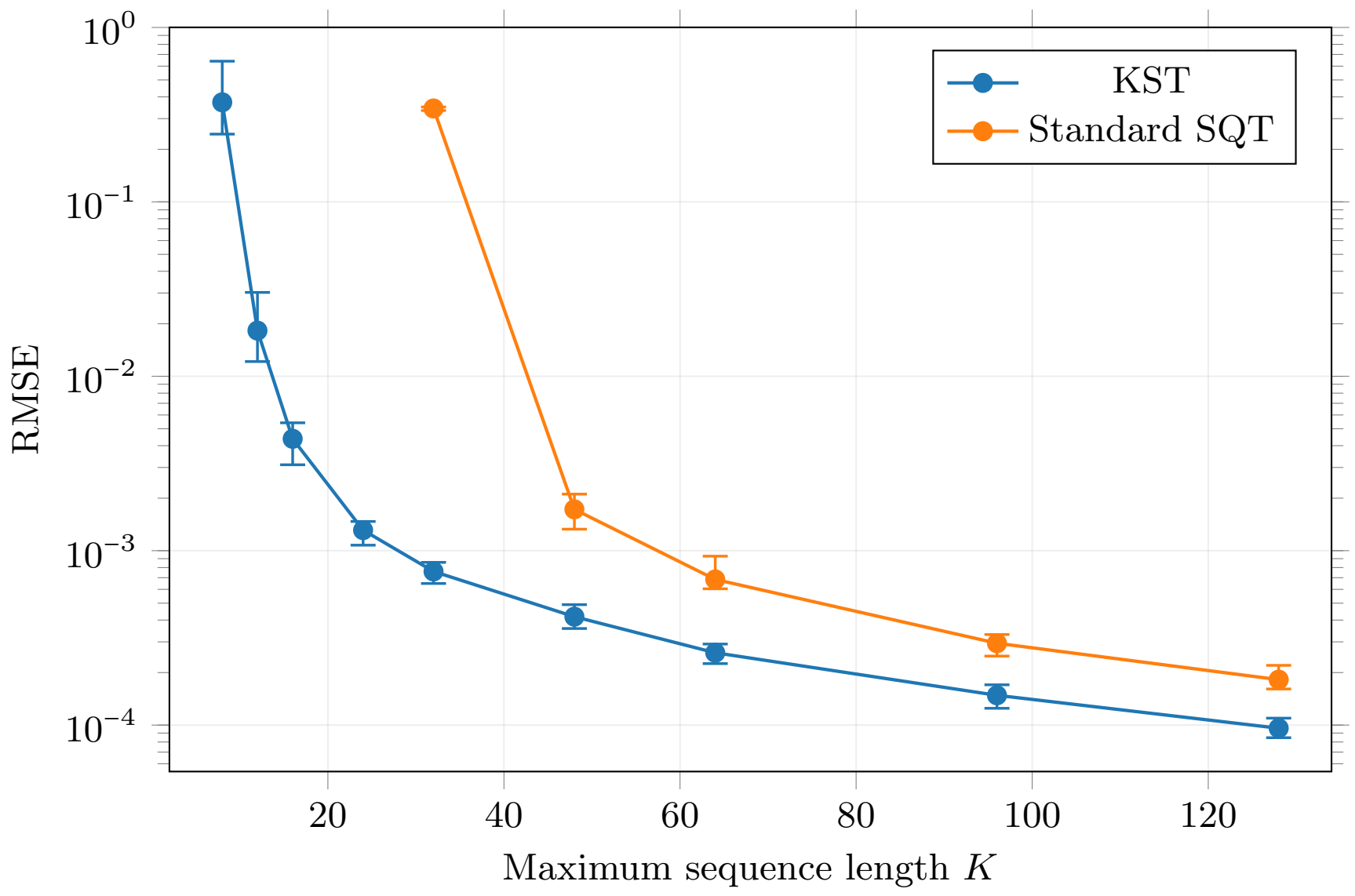

For each maximum sequence length $K$, which takes several values ranging from $8$ to $128$, the experiment contains $100$ independent trials. As shown in Fig.~\ref{fig: pending1}, when $K\leqslant24$, standard SQT yields no acceptable spectral estimate in any of the $100$ trials, so no RMSE is reported. For larger $K$, the median RMSE of KST is substantially smaller than that of SQT. If we instead fix the accuracy threshold at $10^{-3}$, the smallest tested value of $K$ for which the median RMSE falls below this threshold is $32$ for KST and $64$ for SQT.

Because we set $L=R=I_4$ and the target gate is noiseless, both methods use the same measurement data. The only difference is that SQT adds the four signals that KST could handle separately into one. The improvement illustrates the benefit of fitting fewer modes simultaneously: nearby eigenvalues belonging to different sectors can be analyzed independently, easing the separation required in a global SQT fitting.

\subsection{KST across noisy targets}
\label{subsec: sim_noisy_targets}

The second experiment applies KST to $120$ independently sampled noisy two-qubit gates. The Cartan angles are sampled uniformly from $0\leqslant|c_z|\leqslant c_y\leqslant c_x\leqslant\pi/4$, and the local gates $L$ and $R$ are independently Haar sampled and implemented exactly. We implement one trial per setting for each $K\in\{8,16,32,64,96,128\}$.

Reduced-order fitting is used most often in $V_0$, where the three eigenvalues can remain too close to distinguish with the available sequence lengths and number of shots. At $K=128$, this sector is represented by one estimated eigenvalue with multiplicity three in $117$ of the $120$ settings, whereas $348$ of the $360$ fits in the other sectors retain four modes.

\begin{figure}[h!]
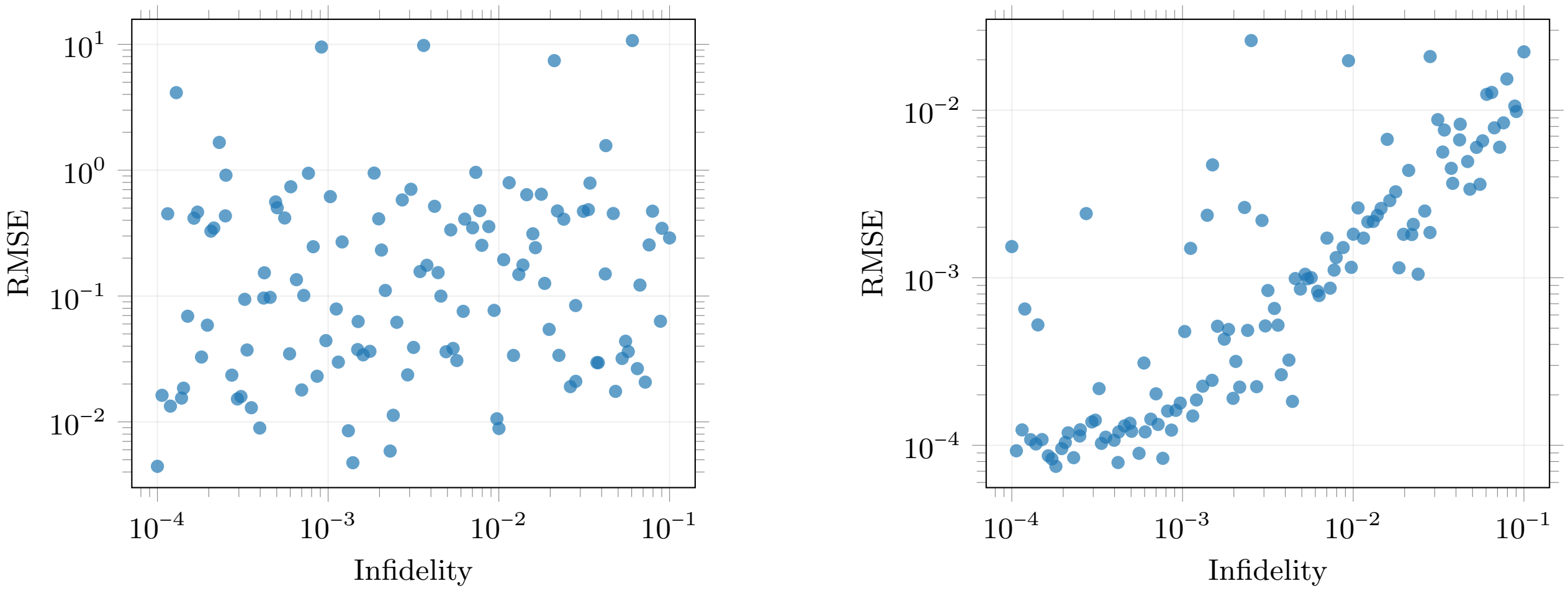

    \centering
    \subfigure[K=8]{
        \begin{minipage}[b]{0.48\linewidth}
            \centering
            \input{figures/pending2_K8_all.tex}
        \end{minipage}
    }
    \hfill
    \subfigure[K=128]{
        \begin{minipage}[b]{0.48\linewidth}
            \centering
            \input{figures/pending2_K128_all.tex}
        \end{minipage}
    }
    \caption{RMSE across the $120$ noisy targets for different $K$. Each point represents one trial for one setting.}
    \label{fig: pending2}
\end{figure}

As shown in Fig.~\ref{fig: pending2}, the median RMSE decreases from $1.24\times10^{-1}$ at $K=8$ to $8.47\times10^{-4}$ at $K=128$. Thus, increasing $K$ substantially improves the typical reconstruction accuracy. At $K=128$, however, the mean RMSE is $2.76\times10^{-3}$, and eight settings still have an RMSE above $10^{-2}$. Accurate reconstruction is therefore achieved for most settings, while the larger errors are concentrated mainly in $V_0$. Closely spaced eigenvalues can remain difficult to resolve when only finite, noisy signals are available. If unresolved eigenvalues are represented by a single common estimate, their contribution to the RMSE is bounded below by a constant, regardless of what the estimate is. Until these eigenvalues are resolved, increasing $K$ cannot reduce this contribution below that lower bound. Conversely, a small RMSE indicates a good approximation of the spectrum but does not imply that every eigenvalue has been resolved individually. Results for all six values of $K$ are shown in Fig.~\ref{fig: full pending2}.



\subsection{SEB and hybrid infidelity estimation}
\label{subsec:sim_hybrid}

The final experiment applies SEB to the same $120$ noisy two-qubit gates, with one trial per setting for each $K\in\{8,16,32,64,96,128\}$. Each of the fifteen Pauli signals is fitted using the first-order matrix-pencil method.

\begin{figure}[h!]
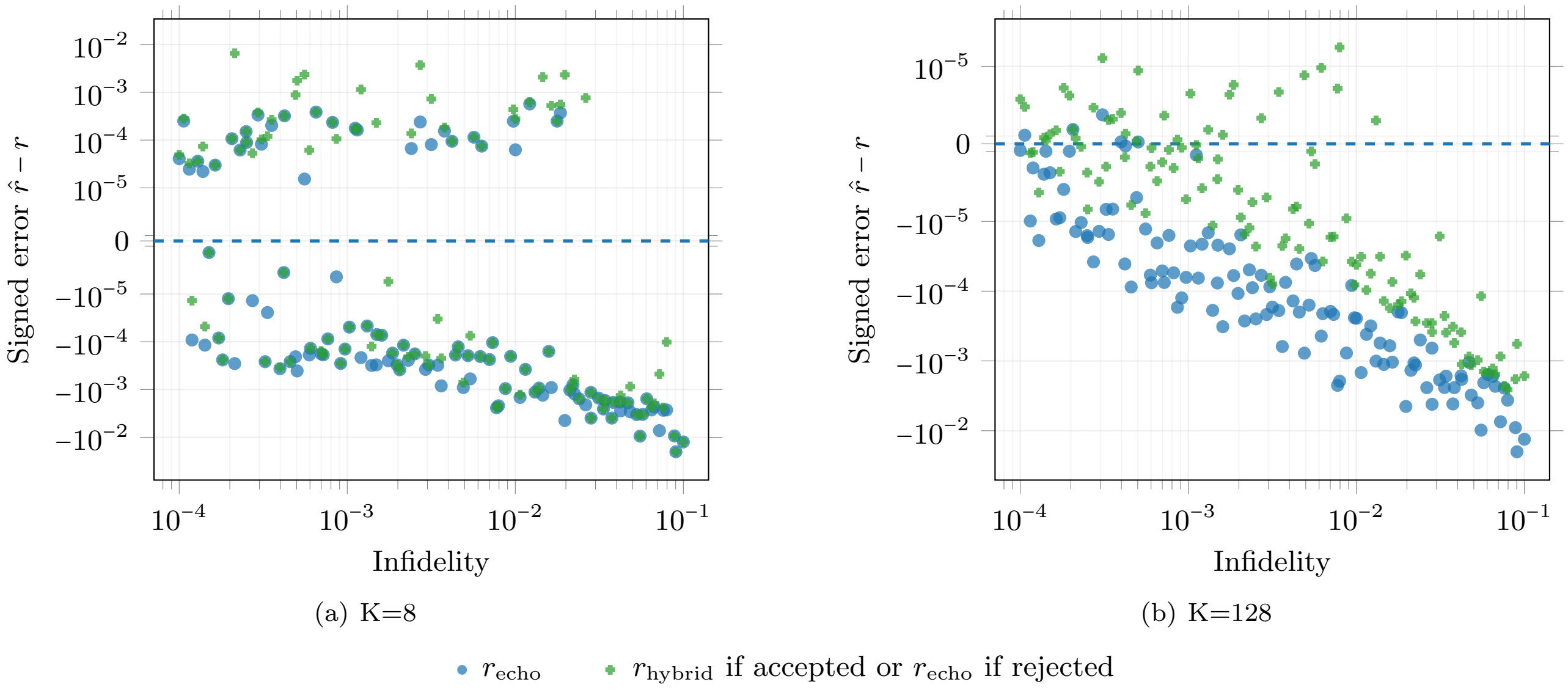

    \centering
    \subfigure[K=8]{
        \begin{minipage}[b]{0.48\linewidth}
            \centering
            \input{figures/bias_K8.tex}
        \end{minipage}
    }
    \hfill
    \subfigure[K=128]{
        \begin{minipage}[b]{0.48\linewidth}
            \centering
            \input{figures/bias_K128.tex}
        \end{minipage}
    }
    \par\vspace{0.3em}
    {\small
    \definecolor{biasBlue}{rgb}{0.12156863,0.46666667,0.70588235}
    \definecolor{biasGreen}{rgb}{0.17254902,0.62745098,0.17254902}
    \tikz[baseline=-0.5ex]{\fill[biasBlue,opacity=0.72] (0,0) circle (1.5pt);}
    \,$r_\textrm{echo}$\qquad
    \tikz[baseline=-0.5ex]{\draw[biasGreen,opacity=0.72,line width=1.5pt] (-1.5pt,0)--(1.5pt,0) (0,-1.5pt)--(0,1.5pt);}
    \,$r_\textrm{hybrid}$ if accepted or $r_\textrm{echo}$ if rejected\par
    }
    \caption{The signed errors of $r_\textrm{echo}$ and $r_\textrm{hybrid}$ versus the true infidelity across the $120$ noisy targets for different $K$. When the Cartan angle fit fails its checks, the Cartan angle correction is abandoned and the reported estimate retains $r_\textrm{echo}$.}
    \label{fig: pending3}
\end{figure}

As shown in Fig.~\ref{fig: pending3}, at $K=128$, the median absolute error of $r_{\textrm{echo}}$ is $1.48\times10^{-4}$, and the mean is $1.10\times10^{-3}$. SEB underestimates the true infidelity in $115$ of the $120$ settings. We then combine each echo estimate with the KST phase information obtained at the same $K$, as described in Sec.~\ref{subsec: hybrid infidelity}. The Cartan angle correction is added only when the phase fit passes the checks; otherwise, the echo estimate is retained. At $K=8$, the phase correction is accepted in $51$ of the $120$ settings, and it does not reduce the median absolute error across the ensemble. At $K=128$, it is accepted in all $120$ settings. The hybrid estimate $r_{\textrm{hybrid}}$ then has a median absolute error of $9.20\times10^{-6}$ and a mean of $2.07\times10^{-4}$, improving on SEB in $111$ of the $120$ settings. At this length, the phase correction therefore substantially improves the infidelity estimate, although the improvement is not uniform across settings. Results for all six values of $K$ are shown in Fig.~\ref{fig: full pending3}.

As $K$ increases, the median absolute error of $r_\textrm{echo}$ changes little, whereas that of $r_\textrm{hybrid}$ continues to decrease. This behavior is consistent with the approximation underlying SEB: longer sequences and more data can improve the precision of the estimates of decay rates, but do not by themselves eliminate the difference between $r$ and $r_{\textrm{echo}}$. The KST phases provide complementary information that can reduce this discrepancy. At short sequence lengths, the phase estimates may not yet be accurate enough for the correction to improve the result.



\section{Comparison with other protocols}\label{sec: comparison}
Like KST, channel spectrum benchmarking (CSB) is also a variant of SQT that reduces the number of modes analyzed in a single signal~\cite{Gu_2023}. It prepares a superposition $\ket{\psi}$ of two eigenstates of the ideal target gate and measures the projector $\ketbra{\psi}$ after repeated applications of the implemented gate. Writing $U\ket{\phi_j}=e^{i\varphi_j}\ket{\phi_j}\ (j=1,2)$, a representative choice is
\begin{equation}
    \begin{gathered}
        \ket{\psi}=\frac{\ket{\phi_1}+\ket{\phi_2}}{\sqrt{2}},\\
        g(m)=\tra{\ketbra{\psi}\tilde{\mathcal{U}}^m(\ketbra{\psi})}.
    \end{gathered}
\end{equation}
With ideal SPAM and noiseless $U$, $g(m)$ involves at most three distinct modes: $1$ and $e^{\pm i(\varphi_1-\varphi_2)}$. SPAM errors and the noise of $\mathcal{U}$ can both introduce other modes into $g(m)$, because a non-ideal $\ketbra{\psi}$ would have overlaps with other eigen-operators, and noise would make $\ketbra{\phi_j}{\phi_k}$ no longer be eigen-operators. To keep this low-order description reliable, one should assume that additional weak modes can be filtered during the matrix-pencil analysis.

KST instead obtains its mode bound through a process similar to character benchmarking~\cite{helsen2019newclassefficientrandomized}. Internal $\mathbb S$-twirling enforces the block diagonal property, and the Pauli character filter isolates each block. The bound of three or four modes therefore does not rely on neglecting small overlaps with eigen-operators in other sectors. This structural distinction does not establish a universal advantage. When the required eigenstate superpositions are easy to prepare, CSB can provide a more flexible probe, and can be extended to multi-qubit situations. KST, in turn, requires only a fixed Pauli preparation and measurement scheme for arbitrary two-qubit targets, but can be affected by errors in auxiliary gates. The use of these two protocols should depend on the specific circumstances and which assumptions are acceptable.

Gate set tomography (GST) provides a detailed, self-consistent characterization of gates and SPAM~\cite{Nielsen_2021}. Recent advances have shown that one can substantially reduce the number of circuits required for two-qubit GST~\cite{rudinger2023twoqubitgatesettomography}. Nevertheless, calibration of physical devices can often be guided by a few gate-quality indicators rather than a complete description of the implemented operations. For such tasks, full tomography may devote experimental and computational resources to parameters that are not needed for the immediate objective. By comparison, our protocols reconstruct $15$ nontrivial eigenvalues and estimate $15$ decay rates, obtaining spectral information and fidelity estimates without reconstructing the entire gate set and SPAM.

\section{Discussion}\label{sec: discussion}
In this work, we have shown how the algebraic structure of quantum gates under local unitary transformations can be exploited to design SPAM-robust characterization protocols beyond standard gate sets. Applying this perspective to arbitrary two-qubit gates via the KAK decomposition, we leveraged the decoupling of the canonical nonlocal core into four invariant sectors and the existence of echo circuits within each sector to develop two complementary protocols, KST and SEB, using only ideal local auxiliary operations. We established the theoretical basis of both protocols and their combination, in which spectral information from KST cancels the leading coherent-error bias in SEB, and demonstrated their performance through numerical simulations. Together, these methods provide practical tools for diagnosing errors and guiding the calibration of native entangling gates beyond the Clifford group.

An important step toward experimental implementation is to relax the assumption of ideal auxiliary local gates. A useful direction for future work is to quantify how imperfections in the local twirling and basis-transformation operations propagate into the reconstructed spectrum and fidelity estimates. Such an analysis would establish the auxiliary-gate accuracy required to achieve a desired precision in the target-gate estimate and clarify when mitigation of auxiliary-gate errors becomes necessary.

Our structural approach also points to natural extensions toward multi-qubit operations. As physical platforms support more complex native gates and larger circuits, it becomes increasingly important to characterize multi-qubit operations as a whole, since errors such as crosstalk during simultaneous gate applications can be missed when components are characterized in isolation~\cite{PhysRevLett.109.240504,PRXQuantum.2.040338,Fan_2025}. A direct class of generalizations consists of multi-qubit gates that reduce under local unitary transformations to evolutions generated by local commuting Pauli Hamiltonians, encompassing many native few-body entangling interactions such as multi-qubit Ising-type and M{\o}lmer--S{\o}rensen-like gates. Because the canonical core of the two-qubit KAK decomposition is itself generated by pairwise commuting Pauli operators, these multi-qubit gates inherit an analogous algebraic structure: the Abelian Pauli subgroup generated by the Hamiltonian terms commutes with the canonical core and defines a local twirling that decouples the operator space into invariant sectors, while local Pauli conjugations that selectively invert generator signs provide a natural mechanism for constructing sector echoes. Extending KST, SEB, and their hybrid correction to such gate families is therefore a promising next step, with practical overhead determined by the resulting sector dimensions and echo depths.

Beyond commuting-generator families, the KAK decomposition of $\mathrm{SU}(4)$~\cite{PhysRevA.67.042313,tucci2005introductioncartanskakdecomposition} admits higher-dimensional analogs for general operations in $\mathrm{SU}(2^n)$, such as recursive Cartan decompositions~\cite{khaneja2000cartandecompositionsu2nconstructive}. A central question for broader multi-qubit benchmarking is to identify which canonical decompositions allow local auxiliary operations to isolate invariant sectors with sufficiently few eigenvalues for spectral reconstruction and to form tractable echoes that cancel the ideal evolution within each sector. Addressing these questions could extend structure-based benchmarking to wider families of multi-qubit gates.

\begin{acknowledgements}
This work was supported by the National Natural Science Foundation of (China Grants No.~12575023), the Innovation Program for Quantum Science and Technology (Grant No.~2021ZD0300804 and No.~2021ZD0300702), and the CCF-QuantumCtek Superconducting Quantum Computing Special Cooperation Program (Grant No.~CCF-QC2025005). GL acknowledges additional funding support from the Turing AI Institute of Nanjing (Grant No.~TR-IIIS 002).
\end{acknowledgements}

\noindent \textbf{AI use statement}: The original ideas and primary proofs were developed by the authors. Generative artificial intelligence tools were used to assist in the derivation of Eq.~\eqref{eq: hybrid accuracy} and the codes used in the numerical simulation. The authors have checked all AI-assisted content and take full responsibility for the manuscript.

\appendix

\section{Properties of the nonlocal part of KAK decomposition}\label{sec: ideal spectrum}

This section details the spectral decomposition of $\mathcal{A}$, where $A$ comes from the KAK decomposition $U=LAR$. In this decomposition, $L$ and $R$ are local unitaries, and $A$ has the following structure
\begin{equation}
    A=\exp[-i(c_xXX+c_yYY+c_zZZ)].
\end{equation}
The Bell states
\begin{equation}
    \ket{\Phi^\pm}=\frac{\ket{00}\pm\ket{11}}{\sqrt2},\ \ket{\Psi^\pm}=\frac{\ket{01}\pm\ket{10}}{\sqrt2},
\end{equation}
form a simultaneous eigenbasis of $XX$, $YY$, and $ZZ$, thus also an eigenbasis of $A$:
\begin{equation}
    \begin{aligned}
        A\ket{\Phi^+}&=e^{-i(c_x-c_y+c_z)}\ket{\Phi^+},\\
        A\ket{\Phi^-}&=e^{-i(-c_x+c_y+c_z)}\ket{\Phi^-},\\
        A\ket{\Psi^+}&=e^{-i(c_x+c_y-c_z)}\ket{\Psi^+},\\
        A\ket{\Psi^-}&=e^{-i(-c_x-c_y-c_z)}\ket{\Psi^-}.
    \end{aligned}
\end{equation}
If $\ket{\psi_1},\ket{\psi_2}$ are two eigenstates of $A$ with eigenvalues $\lambda_1,\lambda_2$ respectively, $\ketbra{\psi_1}{\psi_2}$ is an eigenoperator of $\mathcal{A}$:
\begin{equation}
    \mathcal{A}(\ketbra{\psi_1}{\psi_2})=A\ketbra{\psi_1}{\psi_2}A^\dag=\lambda_1\lambda_2^*\ketbra{\psi_1}{\psi_2}.
\end{equation}
Through this property, one can verify that
\begin{equation}\label{eq: ideal spectrum}
    \begin{gathered}
        \textrm{spec}(\mathcal{A}|_{V_0})=\{1,1,1,1\},\ V_0=\textrm{span}\{\ketbra{\Phi^+}{\Phi^+},\ketbra{\Phi^-}{\Phi^-},\ketbra{\Psi^+}{\Psi^+},\ketbra{\Psi^-}{\Psi^-}\},\\
        \textrm{spec}(\mathcal{A}|_{V_x})=\{e^{\pm2i(c_y+c_z)},e^{\pm2i(c_y-c_z)}\},\ V_x=\textrm{span}\{\ketbra{\Phi^+}{\Psi^+},\ketbra{\Phi^-}{\Psi^-},\ketbra{\Psi^+}{\Phi^+},\ketbra{\Psi^-}{\Phi^-}\},\\
        \textrm{spec}(\mathcal{A}|_{V_y})=\{e^{\pm2i(c_x+c_z)},e^{\pm2i(c_x-c_z)}\},\ V_y=\textrm{span}\{\ketbra{\Phi^+}{\Psi^-},\ketbra{\Phi^-}{\Psi^+},\ketbra{\Psi^+}{\Phi^-},\ketbra{\Psi^-}{\Phi^+}\},\\
        \textrm{spec}(\mathcal{A}|_{V_z})=\{e^{\pm2i(c_x+c_y)},e^{\pm2i(c_x-c_y)}\},\ V_z=\textrm{span}\{\ketbra{\Phi^+}{\Phi^-},\ketbra{\Phi^-}{\Phi^+},\ketbra{\Psi^+}{\Psi^-},\ketbra{\Psi^-}{\Psi^+}\}.
    \end{gathered}
\end{equation}

\section{Details of classical post-processing}\label{sec: classical post processing}
This appendix describes how we assign multiplicities and estimate the Cartan angle errors, as discussed in Secs.~\ref{subsec: mode selection} and~\ref{subsec: hybrid infidelity}, respectively.

\subsection{Multiplicity assignment}\label{subsec: multiplicity}

Suppose that a sector (or the entire traceless operator subspace) with $n$ algebraic modes is represented by $s\leqslant n$ fitted modes as $\hat{g}(m)\approx\sum_{j=1}^{s}a_j\hat{\lambda}_j^m$. We assign multiplicities using the normalized magnitudes of the fitted amplitudes:
\begin{equation}
    \begin{gathered}
        w_j=\frac{n|a_j|}{\sum_{\ell=1}^{s}|a_\ell|},\\
        \boldsymbol{\mu}=\underset{\substack{\mu_j\in\mathbb{N}_+\\ \sum_{j=1}^{s}\mu_j=n}}{\operatorname{arg\,min}}\sum_{j=1}^{s}(\mu_j-w_j)^2.
    \end{gathered}
\end{equation}
Equal multiplicities are also required for complex-conjugate mode pairs. After fixing $\boldsymbol{\mu}$, we repeat each $\hat{\lambda}_j$ exactly $\mu_j$ times and match the expanded list one-to-one to the known ideal eigenvalues by minimizing the total squared distance.

\subsection{Phase estimation}\label{subsec: delta estimation}

Equation~\eqref{eq: ideal spectrum} suggests that Cartan angle errors produce phase shifts of $\pm2(\delta_x\pm\delta_y),\pm2(\delta_x\pm\delta_z)$ and $\pm2(\delta_y\pm\delta_z)$. After matching the estimated eigenvalues to the ideal ones, we obtain linear equations of the form
\begin{equation}\label{eq: phase estimation}
    \begin{pmatrix}
        2&2&0\\
        2&-2&0\\
        \vdots&\vdots&\vdots
    \end{pmatrix}\begin{pmatrix}
        \delta_x\\
        \delta_y\\
        \delta_z
    \end{pmatrix}\approx\begin{pmatrix}
        \arg\hat{\lambda}_1-(2c_x+2c_y)\\
        \arg\hat{\lambda}_2-(2c_x-2c_y)\\
        \vdots
    \end{pmatrix}.
\end{equation}
We solve these equations by linear least squares. It is worth noting that even if we have achieved a matching between the estimated eigenvalues and the ideal eigenvalues, it is still possible that due to the significant difference between them, we consider this matching unreliable, and thus several lines of Eq~\eqref{eq: phase estimation} might be deleted. This Cartan angle estimation is accepted when the phase equations have rank $3$ and the residual RMS is at most $0.02$ rad. Otherwise, the echo estimate is retained and we abandon the inclusion of the Cartan angle correction term.

\section{Derivation of the infidelity estimators}\label{sec: infidelity proof}
In Sec.~\ref{subsec: echo infidelity} and~\ref{subsec: hybrid infidelity}, we gave small error relations between the exact infidelity $r$, the echo estimation $r_\textrm{echo}$, and the hybrid estimation $r_\textrm{hybrid}$, where
\begin{equation}
    \begin{gathered}
        r=1-F(\Lambda),\\
        r_\textrm{echo}=\frac{1}{16}\sum_{P\in V_0}(1-(\mathcal{C}_0)_{PP})+\frac1{32}\sum_{\chi=x,y,z}\sum_{P\in V_\chi}(1-(\mathcal{C}_\chi)_{PP}),\\
        r_\textrm{hybrid}=r_\textrm{echo}+\hat{\delta}_x^2+\hat{\delta}_y^2+\hat{\delta}_z^2.
    \end{gathered}
\end{equation}
This appendix gives full proofs.

\subsection{Echo infidelity}\label{subsec: r echo proof}

This section derives the small-error relation between $r$ and $r_\textrm{echo}$. First note that
\begin{equation}
    \begin{aligned}
        \tilde{\mathcal{A}}_\mathbb{S}&=\mathbb{E}_{P\in\mathbb{S}}\mathcal{P}^\dag\mathcal{A}\Lambda\mathcal{P}\\
        &=\mathbb{E}_{P\in\mathbb{S}}\mathcal{A}\mathcal{P}^\dag\Lambda\mathcal{P}\\
        &=\mathcal{A}\Lambda_\mathbb{S},
    \end{aligned}
\end{equation}
because $A$ commutes with operators in $\mathbb{S}$. For $\chi=x,y,z$, introduce
\begin{equation}
    \begin{gathered}
        \mathcal B_\chi=(\mathcal{J}_\chi\mathcal{A})|_{V_\chi},\\
        \Delta_\chi=I_4-\Lambda_\mathbb{S}|_{V_\chi}.
    \end{gathered}
\end{equation}
Equation~\eqref{eq: echo} implies $\mathcal B_\chi^2=I_4$ because $\mathcal{J}_\chi^\dag=\mathcal{J}_\chi$. Since Pauli twirling preserves diagonal entries, the values obtained from SEB satisfy
\begin{equation}
    \begin{aligned}
        \sum_{P\in V_\chi}(\mathcal{C}_\chi)_{PP}&=\textrm{tr}[(\mathcal{J}_\chi\tilde{\mathcal{A}}_\mathbb{S}\mathcal{J}_\chi\tilde{\mathcal{A}}_\mathbb{S})|_{V_\chi}]\\
        &=\textrm{tr}[\mathcal{B}_\chi(I-\Delta_\chi)\mathcal{B}_\chi(I-\Delta_\chi)]\\
        &=4-2\tra{\Delta_\chi}+\tra{\mathcal B_\chi\Delta_\chi\mathcal B_\chi\Delta_\chi}.
    \end{aligned}
\end{equation}
Using $\tra{\Lambda_{\mathbb S}}=\tra{\Lambda}$ and $\mathcal{A}|_{V_0}=I$, the true infidelity is
\begin{equation}
    r=\frac1{16}\sum_{P\in V_0}(1-(\mathcal{C}_0)_{PP})+\frac1{16}\sum_{\chi=x,y,z}\tra{\Delta_\chi}.
\end{equation}
It follows that
\begin{equation}\label{eq: exact minus echo}
    r=r_\mathrm{echo}+\frac1{32}\sum_{\chi=x,y,z}\tra{\mathcal B_\chi\Delta_\chi\mathcal B_\chi\Delta_\chi}.
\end{equation}
Then the unitarity of $\mathcal B_\chi$ gives
\begin{equation}
    \begin{aligned}
        |r_{\mathrm{echo}}-r|&\leqslant\frac1{32}\sum_{\chi=x,y,z}||\mathcal B_\chi\Delta_\chi\mathcal B_\chi||_F\cdot||\Delta_\chi||_F\\
        &=\frac1{32}\sum_{\chi=x,y,z}||\Delta_\chi||_F^2.
    \end{aligned}
\end{equation}

\subsection{Hybrid correction}\label{subsec: r hybrid proof}

This section derives the small-error relation supporting the hybrid correction. Recall the error model:
\begin{equation}
    \begin{gathered}
        \Lambda=I_{16}+\mathcal L_H+\mathcal D+\mathcal R,\\
        \mathcal{L}_H(O)=-i[H,O],
    \end{gathered}
\end{equation}
where $H$ is the traceless Hermitian generator of the coherent error, $\mathcal{D}$ is the leading dissipative contribution, and $\mathcal{R}$ collects the higher-order terms. We have
\begin{equation}
    \begin{gathered}
        ||H||_F=O(\epsilon),\\
        ||\mathcal{D}||_F=O(\kappa),\\
        ||\mathcal{R}||_F=O(\epsilon^2+\epsilon\kappa+\kappa^2).
    \end{gathered}
\end{equation}
The two small parameters describe coherent and dissipative noise strengths respectively.

For any $P\in\mathbb S$ and operator $O$, we have
\begin{equation}
    (\mathcal{P}^\dag\mathcal{L}_H\mathcal{P})(O)=-iP^\dag[H,POP^\dag]P=-i[P^\dag HP,O].
\end{equation}
Thus
\begin{equation}\label{eq: L conjugate}
    \begin{aligned}
        (\mathcal{L}_H)_\mathbb{S}(O)&=\mathbb{E}_{P\in\mathbb{S}}-i[P^\dag HP,O]\\
        &=-i[H',O]\\
        &=\mathcal{L}_{H'}(O),
    \end{aligned}
\end{equation}
where $H'=\mathbb{E}_{P\in\mathbb{S}}P^\dag HP$. This average removes all the Pauli components of $H$ except $XX,YY,ZZ$. Hence we can write
\begin{equation}
    H'=\delta_xXX+\delta_yYY+\delta_zZZ.
\end{equation}

Define $\mathcal{L}_\chi=\mathcal{L}_{H'}|_{V_\chi}$. By the linearity of twirling,
\begin{equation}
    \begin{gathered}
        \Lambda_\mathbb{S}=I_{16}+\mathcal{L}_{H'}+\mathcal{D}_\mathbb{S}+\mathcal{R}_\mathbb{S},\\
        \Delta_\chi=I_4-\Lambda_\mathbb{S}|_{V_\chi}=-\mathcal{L}_\chi+\mathcal{E}_\chi,
    \end{gathered}
\end{equation}
where
\begin{equation}
    \begin{gathered}
        \mathcal{E}_\chi=-(\mathcal{D}_\mathbb{S}+\mathcal{R}_\mathbb{S})|_{V_\chi},\\
        ||\mathcal{L}_\chi||_F=O(\epsilon),\\
        ||\mathcal{E}_\chi||_F=O(\epsilon^2+\kappa).
    \end{gathered}
\end{equation}

\begin{lemma}\label{lemma: lemma1}
    For $\chi\in\{x,y,z\}$,
    \begin{equation}
        \mathcal{B}_\chi\mathcal{L}_\chi\mathcal{B}_\chi=-\mathcal{L}_\chi.
    \end{equation}
\end{lemma}
\begin{proof}
    Define $P_x=XX,P_y=YY,P_z=ZZ$. Similarly to Eq.~\eqref{eq: L conjugate}, for $\chi\in\{x,y,z\}$, we have
    \begin{equation}
        (\mathcal{J}_\chi\mathcal{A})^\dag\mathcal{L}_{H'}(\mathcal{J}_\chi\mathcal{A})=\mathcal{L}_{(J_\chi A)^\dag H'(J_\chi A)}.
    \end{equation}
    Because $J_\chi$ commutes with $P_\chi$ and anti-commutes with the other two operators, and $A$ commutes with all three operators,
    \begin{equation}
        \begin{aligned}
            A^\dag J_\chi^\dag H' J_\chi A&=A^\dag(-H'+2\delta_\chi P_\chi)A\\
            &=-H'+2\delta_\chi P_\chi.
        \end{aligned}
    \end{equation}
    Any operator $O\in V_\chi$ commutes with $P_\chi$, so $\mathcal{L}_{-H'+2\delta_\chi P_\chi}(O)=\mathcal{L}_{-H'}(O)$. Because $\mathcal{J}_\chi$, $\mathcal{A}$ and $\mathcal{L}_{H'}$ share the same block diagonal structure,
    \begin{equation}
        \begin{aligned}
            \mathcal{B}_\chi\mathcal{L}_\chi\mathcal{B}_\chi&=\mathcal{B}_\chi^\dag\mathcal{L}_\chi\mathcal{B}_\chi\\
            &=[(\mathcal{J}_\chi\mathcal{A})^\dag\mathcal{L}_{H'}(\mathcal{J}_\chi\mathcal{A})]|_{V_\chi}\\
            &=\mathcal{L}_{-H'+2\delta_\chi P_\chi}|_{V_\chi}\\
            &=\mathcal{L}_{-H'}|_{V_\chi}\\
            &=-\mathcal{L}_\chi.
        \end{aligned}
    \end{equation}
\end{proof}

\begin{lemma}\label{lemma: lemma2}
    \begin{equation}
        \sum_{\chi\in\{x,y,z\}}\tra{\mathcal{L}_\chi^2}=-32(\delta_x^2+\delta_y^2+\delta_z^2).
    \end{equation}
\end{lemma}
\begin{proof}
    Because $\mathcal{L}_{H'}$ is block diagonal and its $V_0$ part is zero,
    \begin{equation}
        \begin{aligned}
            \sum_{\chi\in\{x,y,z\}}\tra{\mathcal{L}_\chi^2}&=\tra{\mathcal{L}_{H'}^2}\\
            &=\frac{1}{4}\sum_{P\in\mathbb{P}}\tra{P^\dag\mathcal{L}_{H'}^2(P)}\\
            &=\frac{1}{4}\sum_{P\in\mathbb{P}}\tra{-iP^\dag[H',-i[H',P]]}\\
            &=-\frac{1}{4}\sum_{P\in\mathbb{P}}\tra{P^\dag(H'^2P-2H'PH'+PH'^2)}\\
            &=-8\tra{H'^2}+2\tra{H'}^2\\
            &=-32(\delta_x^2+\delta_y^2+\delta_z^2),
        \end{aligned}
    \end{equation}
    where the second-to-last equality uses the property
    \begin{equation}
        \frac{1}{4}\sum_{P\in\mathbb{P}}P^\dag OP=\tra{O}I.
    \end{equation}
\end{proof}

Using Lemma~\ref{lemma: lemma1} and~\ref{lemma: lemma2}, we have
\begin{equation}
    \begin{aligned}
        \sum_{\chi=x,y,z}\tra{\mathcal B_\chi\Delta_\chi\mathcal B_\chi\Delta_\chi}&=\sum_{\chi=x,y,z}\tra{\mathcal{B}_\chi(-\mathcal{L}_\chi+\mathcal{E}_\chi)\mathcal{B}_\chi(-\mathcal{L}_\chi+\mathcal{E}_\chi)}\\
        &=\sum_{\chi=x,y,z}\tra{-\mathcal{L}_\chi^2+2\mathcal{L}_\chi\mathcal{E}_\chi+\mathcal{B}_\chi\mathcal{E}_\chi\mathcal{B}_\chi\mathcal{E}_\chi}\\
        &=-\sum_{\chi=x,y,z}\tra{\mathcal{L}_\chi^2}+O(\epsilon)\cdot O(\epsilon^2+\kappa)+O((\epsilon^2+\kappa)^2)\\
        &=32(\delta_x^2+\delta_y^2+\delta_z^2)+O(\epsilon^3+\epsilon\kappa+\kappa^2).
    \end{aligned}
\end{equation}

Substituting this expression into Eq.~\eqref{eq: exact minus echo} gives
\begin{equation}
    r=r_\textrm{echo}+\delta_x^2+\delta_y^2+\delta_z^2+O(\epsilon^3+\epsilon\kappa+\kappa^2).
\end{equation}

\section{Details of numerical simulation}\label{sec: detail simulation}

\subsection{Error construction}\label{subsec: error construction}
In the procedure of numerical simulation, the noise channels are constructed by coherent error, amplitude damping, and stochastic Pauli errors. For the coherent component, we draw $15$ independent standard normal variables, normalize the resulting vector, and scale it to obtain the coefficients $h_P$. The resulting channel is
\begin{equation}
    \begin{gathered}
        \Lambda_{\textrm{coh}}(\rho)=e^{-iH_0}\rho e^{iH_0},\\
        H_0=\sum_{P\neq II}h_PP.
    \end{gathered}
\end{equation}
Amplitude damping acts independently on the two qubits. Its single-qubit Kraus operators are
\begin{equation}
    K_0=\begin{pmatrix}
        1&0\\
        0&\sqrt{1-\gamma}
    \end{pmatrix},\ K_1(\gamma)=\begin{pmatrix}
        0&\sqrt{\gamma}\\
        0&0
    \end{pmatrix}.
\end{equation}
The stochastic component is a Pauli channel, described by
\begin{equation}
    \begin{gathered}
        \Lambda_{\textrm{sto}}(\rho)=\sum_{P\in\mathbb{P}}\lambda_PP\rho P^\dag,\\
        \sum_{P\in\mathbb{P}}\lambda_P=1,\ \lambda_P\geqslant0.
    \end{gathered}
\end{equation}
The complete noise channel is the composition
\begin{equation}
    \Lambda=\Lambda_{\textrm{sto}}\Lambda_{\textrm{damp}}\Lambda_{\textrm{coh}}.
\end{equation}
By adjusting the above parameters, we construct a noise channel ensemble whose infidelities range from $10^{-4}$ to $10^{-1}$. More specifically, we first sample the noise parameters and then adjust a common parameter controlling the three error components until the process infidelity reaches the desired value.

The preparation error acts after the ideal preparation and the measurement error acts before the ideal measurement. On each qubit, a coherent rotation is followed by amplitude damping and single-qubit depolarization. The preparation and measurement noise channels are tensor products of these local compositions.

\clearpage
\subsection{Complete results of simulation}\label{subsec: full figure}

\begin{figure}[h!]
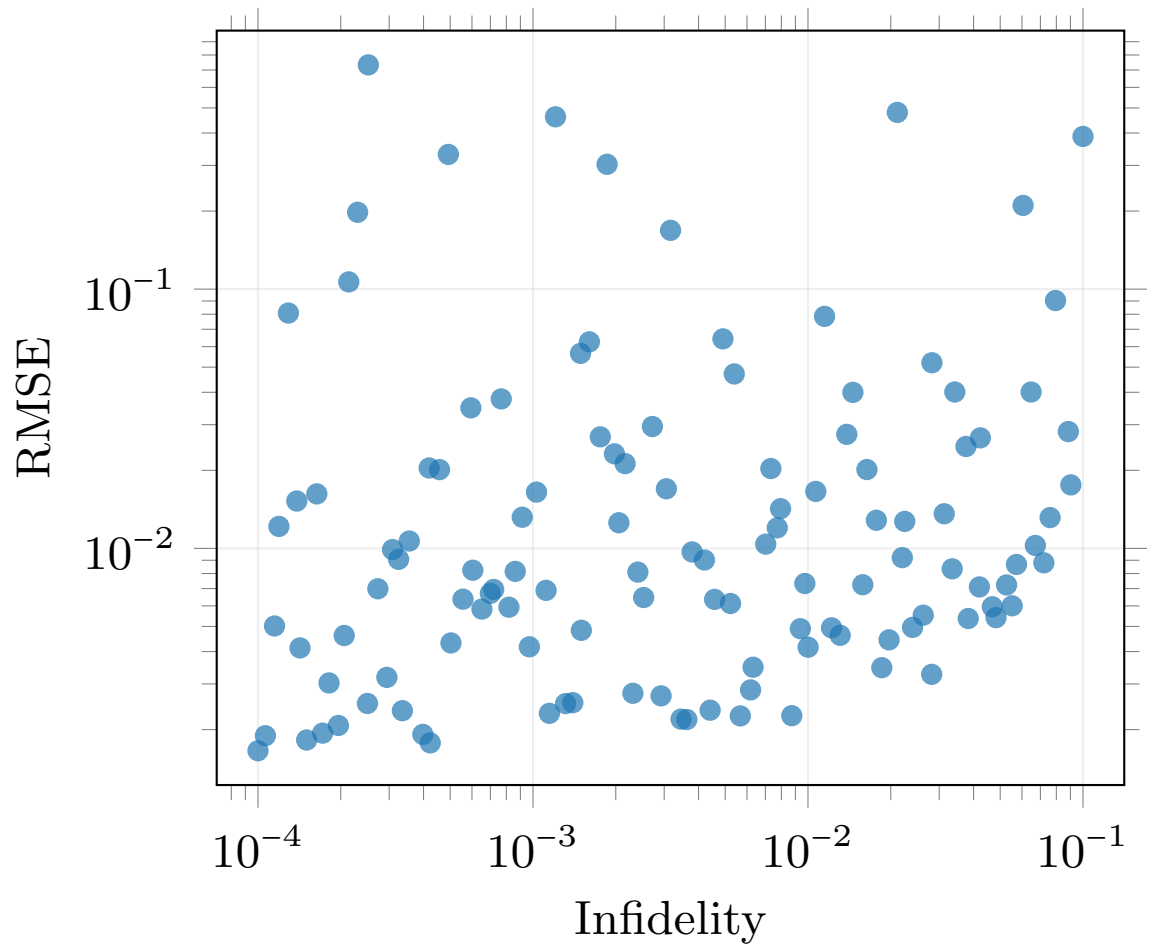
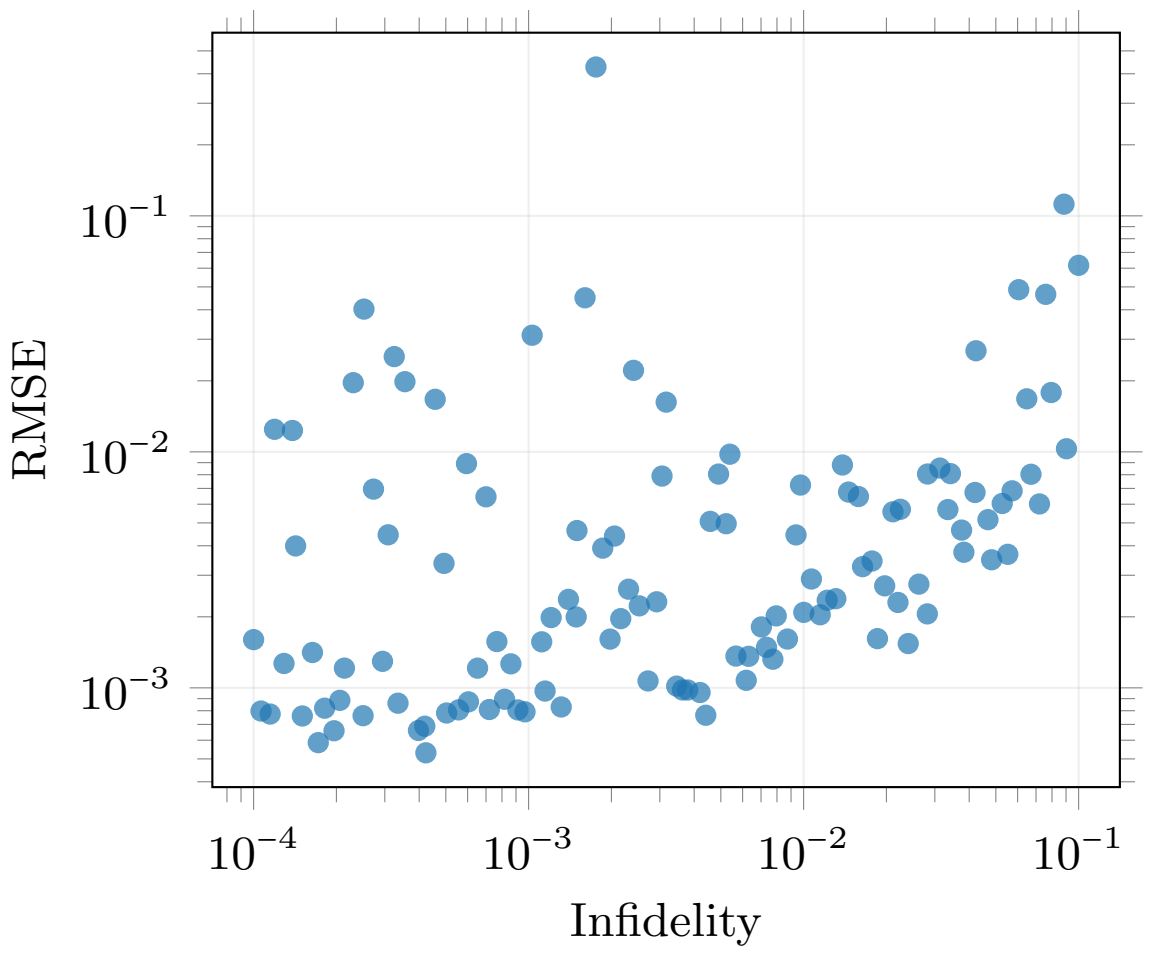
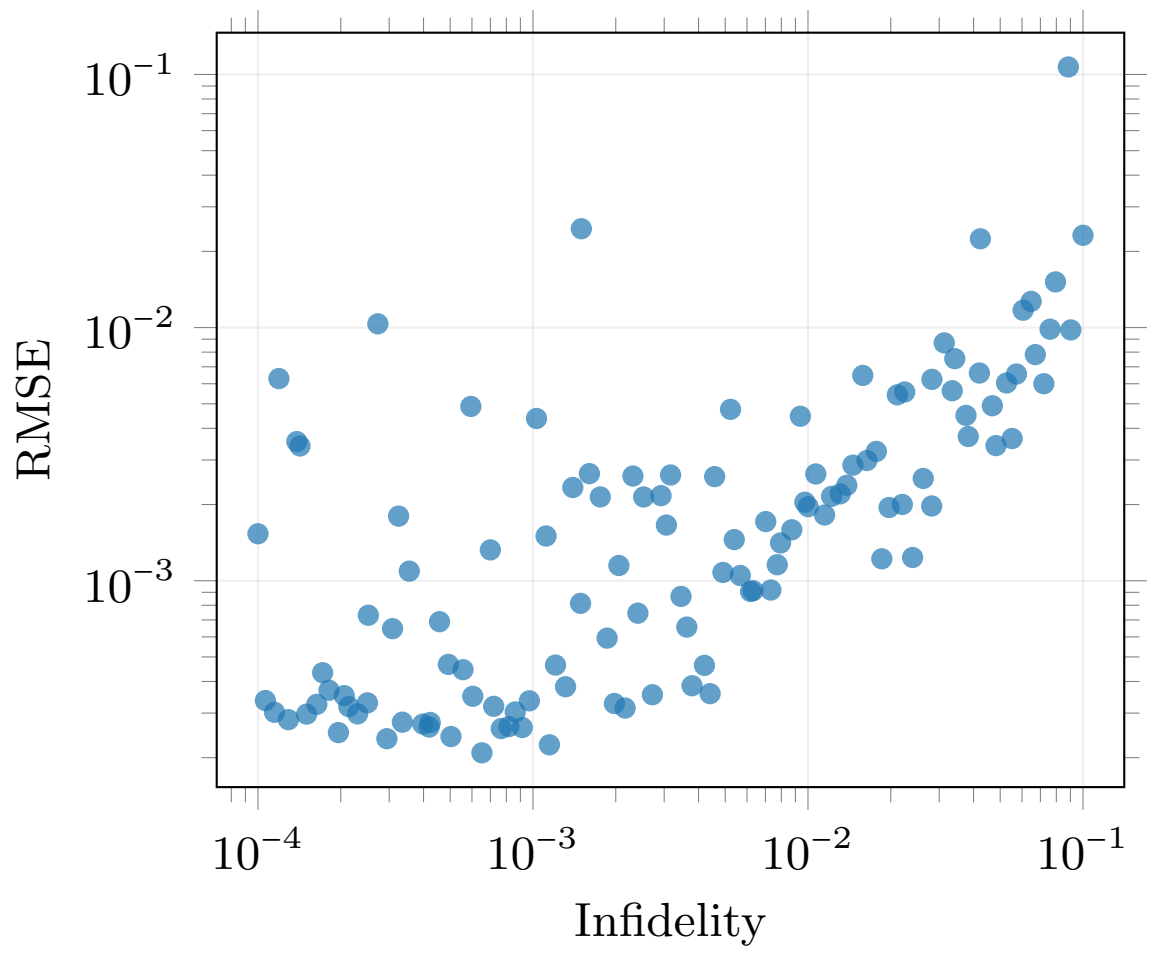
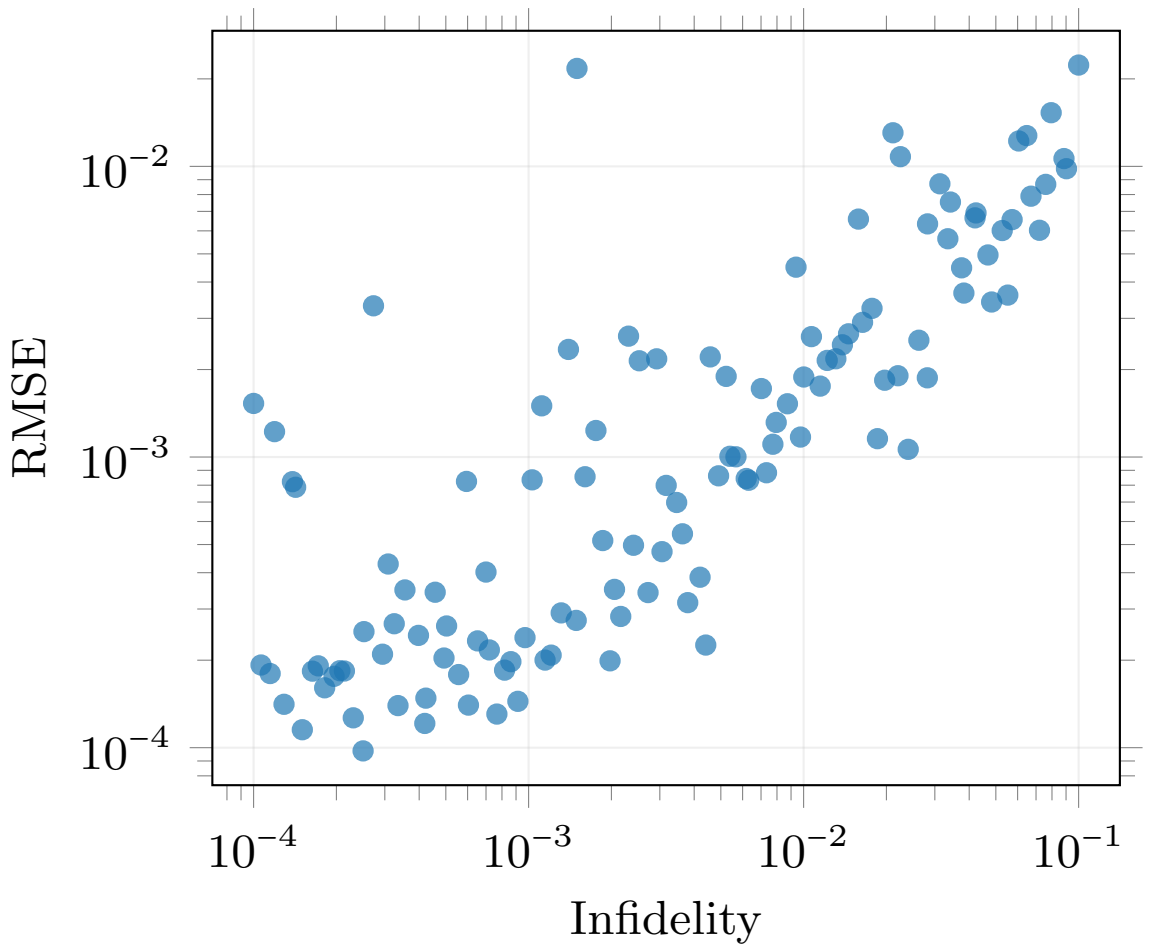
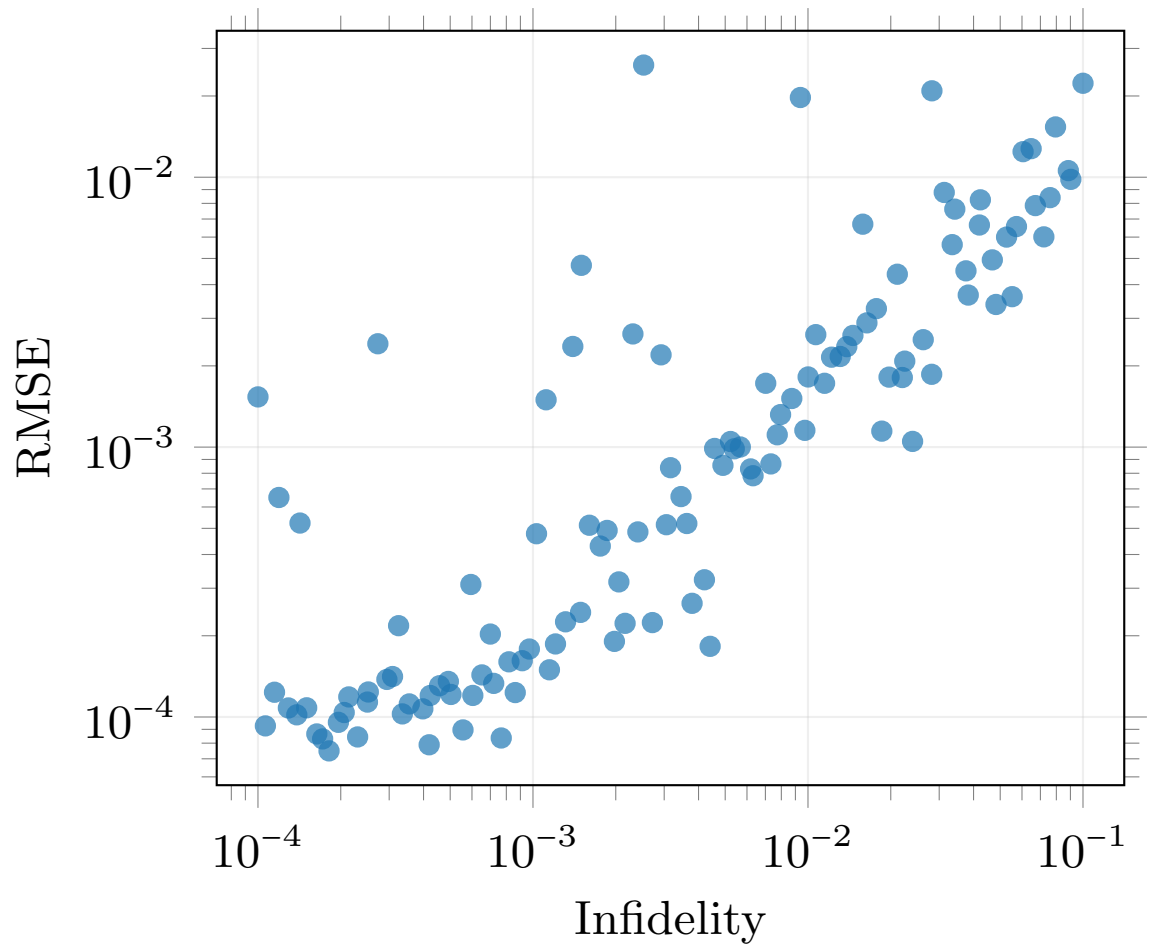

    \centering
    \subfigure[K=8]{
        \begin{minipage}[b]{0.48\linewidth}
            \centering
            \input{figures/pending2_K8_all.tex}
        \end{minipage}
    }
    \hfill
    \subfigure[K=16]{
        \begin{minipage}[b]{0.48\linewidth}
            \centering
            \input{figures/pending2_K16_all.tex}
        \end{minipage}
    }\\
    \subfigure[K=32]{
        \begin{minipage}[b]{0.48\linewidth}
            \centering
            \input{figures/pending2_K32_all.tex}
        \end{minipage}
    }
    \hfill
    \subfigure[K=64]{
        \begin{minipage}[b]{0.48\linewidth}
            \centering
            \input{figures/pending2_K64_all.tex}
        \end{minipage}
    }\\
    \subfigure[K=96]{
        \begin{minipage}[b]{0.48\linewidth}
            \centering
            \input{figures/pending2_K96_all.tex}
        \end{minipage}
    }
    \hfill
    \subfigure[K=128]{
        \begin{minipage}[b]{0.48\linewidth}
            \centering
            \input{figures/pending2_K128_all.tex}
        \end{minipage}
    }\\
    \caption{The full results of the simulation of KST. Each point represents one trial for one setting.}
    \label{fig: full pending2}
\end{figure}

\begin{figure}[h!]
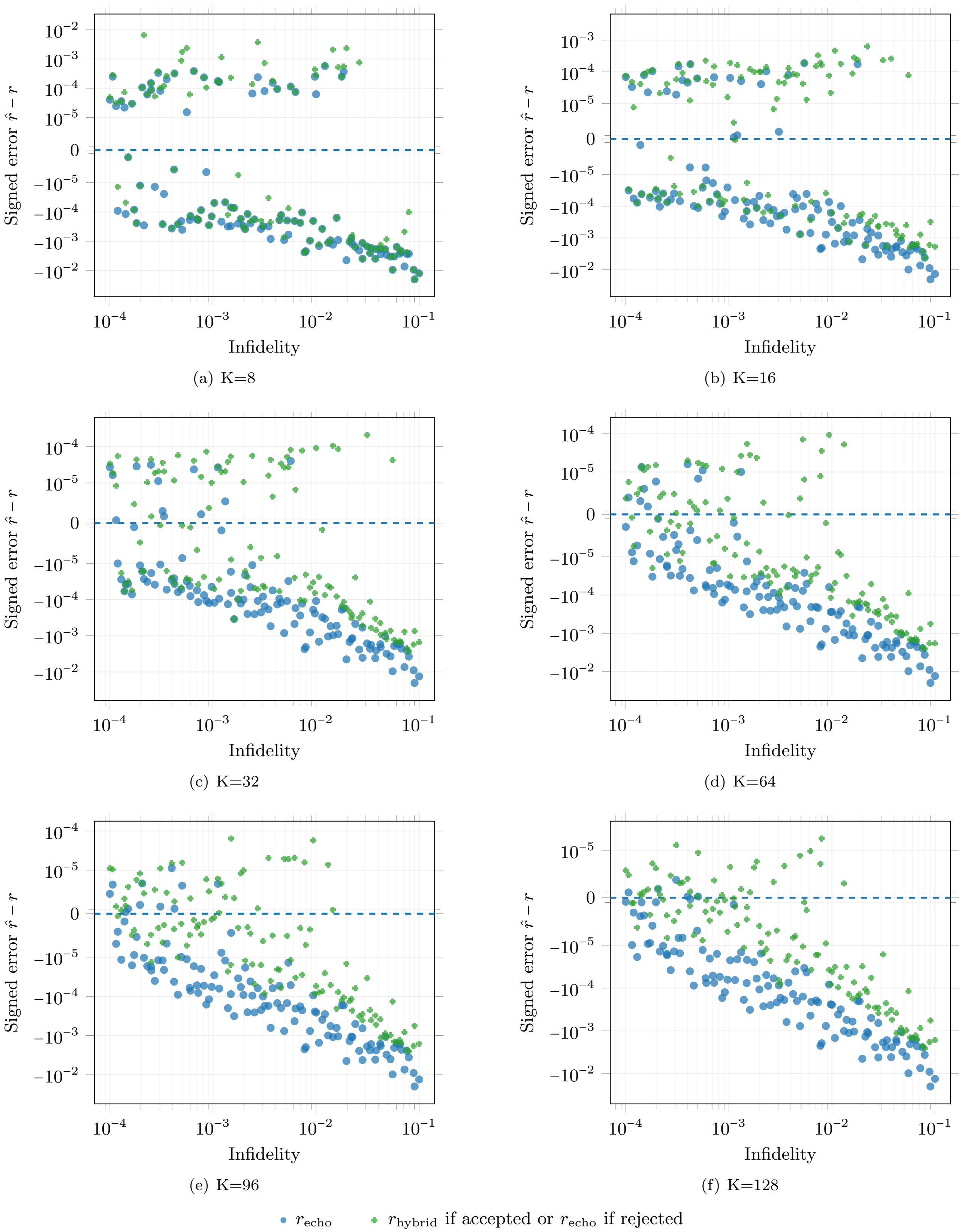

    \centering
    \subfigure[K=8]{
        \begin{minipage}[b]{0.48\linewidth}
            \centering
            \input{figures/bias_K8.tex}
        \end{minipage}
    }
    \hfill
    \subfigure[K=16]{
        \begin{minipage}[b]{0.48\linewidth}
            \centering
            \input{figures/bias_K16.tex}
        \end{minipage}
    }\\
    \subfigure[K=32]{
        \begin{minipage}[b]{0.48\linewidth}
            \centering
            \input{figures/bias_K32.tex}
        \end{minipage}
    }
    \hfill
    \subfigure[K=64]{
        \begin{minipage}[b]{0.48\linewidth}
            \centering
            \input{figures/bias_K64.tex}
        \end{minipage}
    }\\
    \subfigure[K=96]{
        \begin{minipage}[b]{0.48\linewidth}
            \centering
            \input{figures/bias_K96.tex}
        \end{minipage}
    }
    \hfill
    \subfigure[K=128]{
        \begin{minipage}[b]{0.48\linewidth}
            \centering
            \input{figures/bias_K128.tex}
        \end{minipage}
    }\\
    \par\vspace{0.3em}
    {\small
    \definecolor{biasBlue}{rgb}{0.12156863,0.46666667,0.70588235}
    \definecolor{biasGreen}{rgb}{0.17254902,0.62745098,0.17254902}
    \tikz[baseline=-0.5ex]{\fill[biasBlue,opacity=0.72] (0,0) circle (1.5pt);}
    \,$r_\textrm{echo}$\qquad
    \tikz[baseline=-0.5ex]{\draw[biasGreen,opacity=0.72,line width=1.5pt] (-1.5pt,0)--(1.5pt,0) (0,-1.5pt)--(0,1.5pt);}
    \,$r_\textrm{hybrid}$ if accepted or $r_\textrm{echo}$ if rejected\par
    }
    \caption{The full results of the simulation of SEB and the hybrid protocol.}
    \label{fig: full pending3}
\end{figure}


\bibliography{bibtwoqubit}

\end{document}